\documentclass[11pt]{article}
\usepackage{natbib,graphicx,amsmath,amssymb,fullpage}
\newtheorem{theorem}{Theorem}
\newtheorem{lemma}{Lemma}
\newtheorem{definition}{Definition}

\newcommand{\mb}{\mathbf}
\newcommand{\mbb}{\boldsymbol}

\newcommand{\iid}{i.i.d.\ }
\newcommand{\lhs}{l.h.s.\ }
\newcommand{\rhs}{r.h.s.\ }

\title{Group-bound: confidence intervals for groups of variables in sparse high-dimensional regression without assumptions on the design}
\author{Nicolai Meinshausen \\ Seminar f\"ur Statistik, ETH Z\"urich \\ meinshausen@stat.math.ethz.ch}

\begin{document}
\maketitle

\begin{abstract}
It is in general challenging to provide confidence intervals for individual variables in high-dimensional regression without making strict or unverifiable assumptions on the design matrix. We show here that a ``group-bound'' confidence interval can be derived without making \emph{any} assumptions on the design matrix.  The lower bound for the regression coefficient of individual variables can be derived via linear programming. The idea also generalises naturally to groups of variables, where we can derive a one-sided confidence interval for the joint effect of a group. While the confidence intervals of individual variables are by the nature of the problem often very wide, it is shown   to be possible to detect the contribution of  groups of highly correlated predictor variables even when no variable individually shows a significant effect. The assumptions necessary to detect the effect of groups of variables are  shown to be weaker than the weakest known assumptions to detect the effect of individual variables.  
\end{abstract}

\section{Introduction}

High-dimensional linear models have been studied extensively in the last years. The $\ell_1$-penalised Lasso-estimator \citep{tibshirani96regression} has received a majority of the attention, partially due to pairing attractive computational properties with variable selection. The properties of the Lasso estimator have been studied among many other works in a series of papers including \citet{greenshtein03persistency}, \citet{zhang06model} and \citet{bickel07dantzig}. For a good overview see \citet{buhlmann2011statistics}. Computational algorithms include \citet{osborne00on} and \citet{efron04least}. 

To fix notation, assume we have a random response vector $\mb Y\in\mathbb{R}^n$ with expected value $\mathbb{E}(\mb Y)$ and a design matrix $\mb X\in\mathbb{R}^{n\times p}$ (vectors and matrices are shown in boldface throughout). Let $\mbb \beta^* \in\mathbb{R}^p$ be the $\ell_1$-sparsest \emph{Basis Pursuit} \citep{chen01atomic} solution to the noise-free problem \begin{equation} \label{bp} \mbb \beta^* \; =\; \mbox{argmin}_{\mbb \beta\in\mathbb{R}^p} \;\|\mbb \beta\|_1 \;\;\mbox{such that } \mathbb{E}(\mb Y) = \mb X\mbb \beta.\end{equation} While we assume that there is at least a single solution for the latter equality for $p>n$, we take $\mbb \beta^*$ to be an arbitrary member of the set of solutions if the solution is not unique in~(\ref{bp}). 
 The observations $\mb Y$ are now corrupted by some noise $\mbb \varepsilon$ so that $\mb Y=\mb X\mbb \beta^*+\mbb \varepsilon$, which is drawn \iid from a noise distribution with  known distributional form but unknown noise level. If we are not interested in the $\ell_1$-sparsest regression vector~(\ref{bp}) but, for example, the $\ell_0$-sparsest vector then we do need weak assumptions to show equivalence between the solutions, namely the \emph{nullspace condition} in the case of the $\ell_0$-sparsest solution which is discussed in  Section~\ref{section:assumptions}. For the following, however, we will work on inference about the $\ell_1$-sparsest optimal regression vector~(\ref{bp}) and will try to produce confidence intervals with correct coverage that are valid for all designs. 

Statistical inference about $\mbb\beta^*$ in terms of significance tests and confidence intervals for the solution in~(\ref{bp}) has only recently received substantial attention. While the overall stability of estimated sparse solutions was sometimes analysed and exploited for better structure discovery \citep{meinshausen2008ss,shah2013variable,lim2013estimation}, formal significance tests were provided 
in \citet{wasserman2009high} and \citet{meinshausen09pvalues}. They relied on sample splitting of the data. On one half of the data, the Lasso or a similar sparse estimation procedure selects a model which has to be assumed to include the true set of non-zero coefficients in $\mbb \beta^*$ with high probability. The small set of selected variables can then be formally tested with traditional tests on the second half of the data. An issue with this approach is that its validity relies in general on a so-called \emph{beta-min} condition. The condition requires that the smallest non-zero value of $\mbb \beta^*$ is bounded away from zero by a potentially non-negligible amount. 
In contrast, \citet{lockhart2012significance} derived a test for variables along the Lasso solution path. For each variable that enters the model one can test  whether the new variable is significant, conditional on all important variables being included in the model. 
An alternative approach for unconditional confidence intervals uses the fact that the optimal regression coefficient can also be expressed as being proportional to the cross-product of the  residuals of a variable and the response, where the residual is with respect to a regression on all other variables. This has been exploited in an interesting way in  \citet{zhang2011confidence,van2013asymptotically} and \citet{javanmard2013confidence}.  These  approaches rely typically on specific assumptions about the design, the  \emph{compatibility condition} \citep{van2009conditions} being the weakest assumption. It is, however, still a strong condition and often violated in practice due to high correlation between variables. 

Here, we propose a confidence interval (and related test) that provides valid error control  without making \emph{any} assumptions about the design matrix.
 The approach also extends naturally to groups of variables $G\subseteq\{1,\ldots,p\}$ and can provide error confidence intervals for the norms $\|\mbb \beta^*_G\|_q$ for any $q\ge 1$, using only convex optimisation or linear programming in the specific case of $q=1$. Likewise, tests of the null hypothesis $H_{0,G} : \mbb \beta^*_G \equiv \mbb 0 $ can be performed, where $\mbb \beta_G\in \mathbb{R}^{|G|}$ is the vector of coefficients of variables in group $G$. Grouping of variables in high-dimensional regression is natural and some estimators exploit a group structure 
\citep{yuan05msa,meier06group}. While sup-norm bounds on the coefficients as in \citet{lounici2008snc}  can be used to construct confidence intervals for groups of variables, the proposed procedure is to the best of our knowledge the first to combine the following properties:
\begin{enumerate}
\item[(a)] The confidence intervals are valid under \emph{any} design matrix $\mb X$ even in the high-dimensional case as long as are interested in the $\ell_1$-sparsest regression vector~(\ref{bp}). 
\item[(b)] The test has a hierarchical monotonicity property in that if we can reject $H_{0,G}:\mbb \beta_G\equiv \mbb 0$ at some level for a group of variables $G\subseteq\{1,\ldots,p\}$, then the test will also reject $H_{0,G'}$ at the same level if $G\subseteq G'$. Furthermore, the test is adjusted for multiplicity and the level is valid simultaneously for all possible subsets of variables $G\subseteq\{1,\ldots,p\}$.
\item[(c)] The power of the test is not affected by high or perfect correlation between variables in the same group. If we can reject $H_{0,G}$, then the test will also reject $H_{0,G}$ if we add a copy of a variable in $G$ to the design and include it in the group~$G$. We show that the design conditions needed to detect interesting groups of variables are substantially weaker than the conditions needed to detect individually important variables with other approaches. 
\end{enumerate}
The tests rely, though, on knowledge of the distributional form of the error term. To keep the exposition as simple as possible, we will assume that error are rotationally invariant and most examples are provided for Gaussian noise with  unknown noise level, but extensions to more heavy-tailed error distributions are possible. The construction of the confidence interval is proposed and shown to provide valid error control in Section~\ref{section:2}. Some empirical results are shown in Section~\ref{section:numerical}, before concluding with a brief discussion in Section~\ref{section:discussion}.

\section{Confidence intervals for groups of variables}\label{section:2}
Suppose $G\subseteq\{1,\ldots,p\}$ is a group of variables and we want to have a one-sided confidence interval for the $\ell_q$-norm of the coefficients in the group, $\|\mbb \beta^*_G\|_q$, for some $q\ge 1$ or a test for the joint effect of the group, $\| \mb X_G \mbb \beta^*_G\|_2$.
The groups can correspond to individual variables, but the desire to test  group of multiple variables  arises naturally for highly correlated designs. Each individual variable is unlikely to be significant since its effect can typically be explained by some other highly correlated variable. However, when grouping highly correlated variables, we are often able to detect a joint group effect even if we are unable to say \emph{which} variables in the group are responsible.

Any construction of a test for the null hypothesis 
\[ H_{0,G}: \; \mbb \beta^*_G\equiv \mbb 0\]
has to rest on the fact that $\mbb \beta^*$ is the sparsest approximation of $\mb X\mbb \beta=\mathbb{E}(\mb Y)$. We will work with the $\ell_1$-norm ($q=1$) but similar constructions are possible for $q\ge 2$. Define the Basis Pursuit solution \citep{chen01atomic}  as $b(\mb X,\mb Y): \mathbb{R}^n \mapsto \mathbb{R}^p$,
\begin{equation}\label{def:b} b(\mb X,\mb Y) = \mbox{ argmin}_{\mbb \beta\in\mathbb{R}^p} \; \| \mbb \beta\|_1 \;\mbox{ such that } \mb X\mbb \beta=\mb Y .\end{equation}
If the solution is not unique, an arbitrary member of the set of solutions is returned. 
Now, we know that $\mbb \beta^*$ is by definition the Basis Pursuit solution for the noise-free signal $\mathbb{E}(\mb Y)$, that is $\mbb \beta^*=b(\mb X,\mb Y-\mbb \varepsilon)$ and we will exploit this in the following. 
Specifically,  let $C_\alpha\subseteq \mathbb{R}^{p+n}$ for some set $N_\alpha\subseteq \mathbb{R}^n$  be defined as  
\begin{equation}\label{def:C} C_\alpha \; :=\; \big\{  (\mbb \beta,\mbb \eta) \in (\mathbb{R}^p,\mathbb{R}^n)  \;  \big|\;   \mbb \eta \in N_\alpha  \mbox{ and }  \mbb \beta = b(\mb X,\mb Y+\mbb \eta)     \big\},\end{equation}
where the (possibly random) set $N_\alpha$ has to fulfil $\mathbb{P}(-\mbb \varepsilon \in N_\alpha)\ge 1-\alpha$. Using such a set, 
define the lower ``group-bound''  of the one-sided confidence interval for $\|\mbb \beta_G\|_1$  as 
\begin{equation}\label{def:stat}  T_G\; :=\; \mbox{min}_{(\mbb \beta,\mbb \eta)\in C_\alpha}  \|\mbb \beta_G\|_1    .\end{equation}
We suppress  the dependence on $\alpha$  in $T_G$ for notational simplicity.
We can then show correct coverage in the following sense. 
\begin{theorem}\label{th:1}
The one-sided interval $[T_G,\infty)$ is a valid $1-\alpha$ confidence interval for $\|\mbb \beta^*_G\|_1$, simultaneously valid for all subsets of variables $G\subseteq\{1,\ldots,p\}$,
\[ \mathbb{P}\Big( \forall G\subseteq\{1,\ldots,p\}: \; T_G \le \|\mbb \beta^*_G\|_1\Big) \; \ge\;  1-\alpha.\]
\end{theorem}

\emph{Proof: }
The proof follows very directly. The event $-\mbb \varepsilon \in N_\alpha$ is equivalent to the event $(\mbb \beta^*,-\mbb \varepsilon) \in C_\alpha$  since $\mbb \beta^*= b(\mb X,\mb Y-\mbb \varepsilon)=b(\mb X,\mathbb{E}(\mb Y))$.  The event $(\mbb \beta^*,-\mbb \varepsilon) \in C_\alpha$ on the other hand implies that $T_G\le \|\mbb \beta^*_G\|_1$ for all subsets $G\subseteq\{1,\ldots,p\}$ by construction of the statistics~(\ref{def:stat}). Hence,
\begin{equation}\label{res:alpha}  \mathbb{P}\Big( \forall G\subseteq\{1,\ldots,p\}: \; T_G \le \|\mbb \beta^*_G\|_1\Big) \;\ge\; \mathbb{P}\Big( (\mbb \beta^*,-\mbb \varepsilon) \in C_\alpha\Big) \;=\; \mathbb{P}(-\mbb\varepsilon\in N_\alpha) \; \ge \; 1-\alpha  \end{equation}
and $[T_G,\infty)$ is a valid $1-\alpha$ confidence interval for $\|\mbb \beta^*_G\|_1$, simultaneously for all subsets of the variables. \hfill $\square$
\paragraph{}
An immediate consequence is that  the null hypothesis $H_{0,G}:\mbb \beta^*_G\equiv \mb 0$ can be rejected at level $\alpha$ for all groups $G$ for which $T_G>0$ and the probability of  erroneously rejecting a group is bounded by the chosen level. No adjustment for multiplicity is necessary. The type I error is controlled simultaneously for all groups at the chosen level.

The problem with the estimator is that the optimisation in~(\ref{def:stat}) with feasible region $C_\alpha$ of~(\ref{def:C}) can be cumbersome if $C_\alpha$ is not a convex set.  We will strive to find a tight convex relaxation of $C_\alpha$ in~(\ref{def:C}). The results of Theorem \ref{th:1} are clearly still valid if we use a set $\bar C_\alpha $ for which $C_\alpha\subseteq \bar C_\alpha$. However, if the set $\bar C_\alpha$ is very large, the method will become unduly conservative. 
To take an example, we could take $N_\alpha$ to be an appropriate $\ell_2$-ball and replace the constraint $\mbb \beta=b(\mb X,\mb Y+\mbb \eta)$ with the linear constraint $\mb Y+\mbb \eta=\mb X\mbb \beta$. Then $(\mbb \beta,\mbb \eta)\in C_\alpha$ would be identical to a $\ell_2$-constraint on the residuals, $\|\mb Y-\mb X\mbb \beta\|_2\le \lambda$ for some $\lambda>0$. The minimum in~(\ref{def:stat}) would then be 0 for most high-dimensional designs with $p>n$ as the effect of variables in a group $G$ could always identically be replicated by some variables in $\{1,\ldots,p\}\setminus G$.

 We will thus aim to find a convex set $N_\alpha$ for which $\mathbb{P}(-\mbb \varepsilon \in N_\alpha)\ge 1-\alpha$  is exact (or the bound is very tight) and  a close convex approximation to~(\ref{def:C}). Rather than using an $\ell_2$-ball for $N_\alpha$, it will turn out to be computationally attractive to use the convex hull of a number of randomly sampled points on an $\ell_2$-sphere.  The Basis Pursuit constraint in~(\ref{def:C}) can then easily be relaxed to yield a convex set $\bar C_\alpha$, for which efficient optimisation routines are available.

\subsection{Convex hull constraint}
We use as constraint $N_\alpha$ in~(\ref{def:C}) a convex hull of a finite number of points. 
 Let $\mbb e^{(1)},\ldots,\mbb e^{(m)}$ be $m$ samples from a rotationally-invariant distribution in $\mathbb{R}^n$
and rescaled such that $\|\mbb e^{(j)}\|_2=1$ for all $1\le j\le m$. We can for example sample from a standard Gaussian distribution but the results are identical under any other rotationally invariant distribution.
Let $\mb E\in\mathbb{R}^{n\times 2m}$ be the matrix with columns 
\begin{equation}\label{def:M} \mb E_{\cdot j} =\left\{ \begin{array}{cl}  \mbb e^{(\lceil j/2\rceil )} & \mbox{ if } j \mbox{ odd} \\ -\mbb e^{(\lceil j/2\rceil )} & \mbox{ if } j \mbox{ even}  \end{array}  \right. \end{equation}
\begin{figure}
\includegraphics[width=0.45\textwidth]{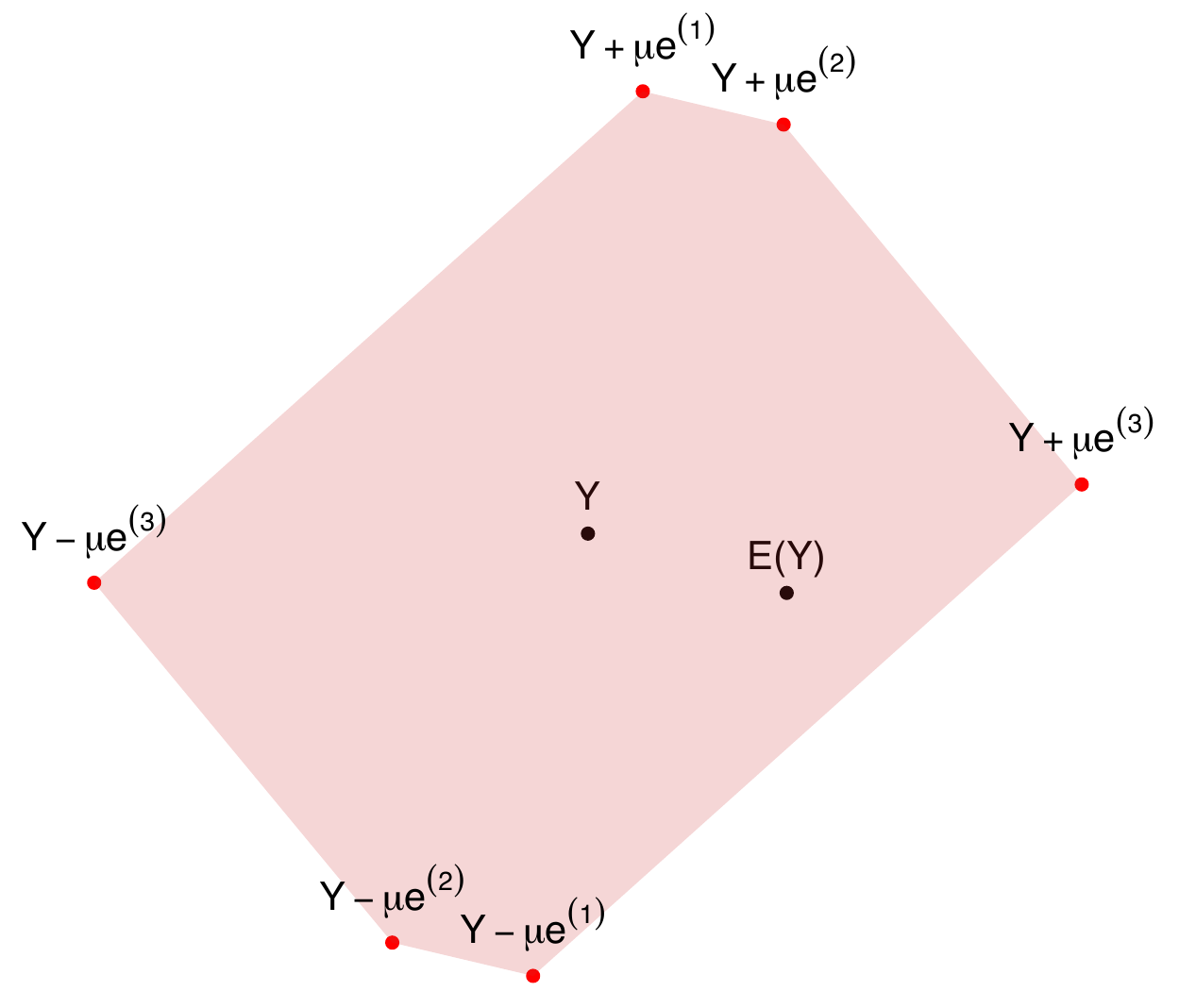}
\includegraphics[width=0.45\textwidth]{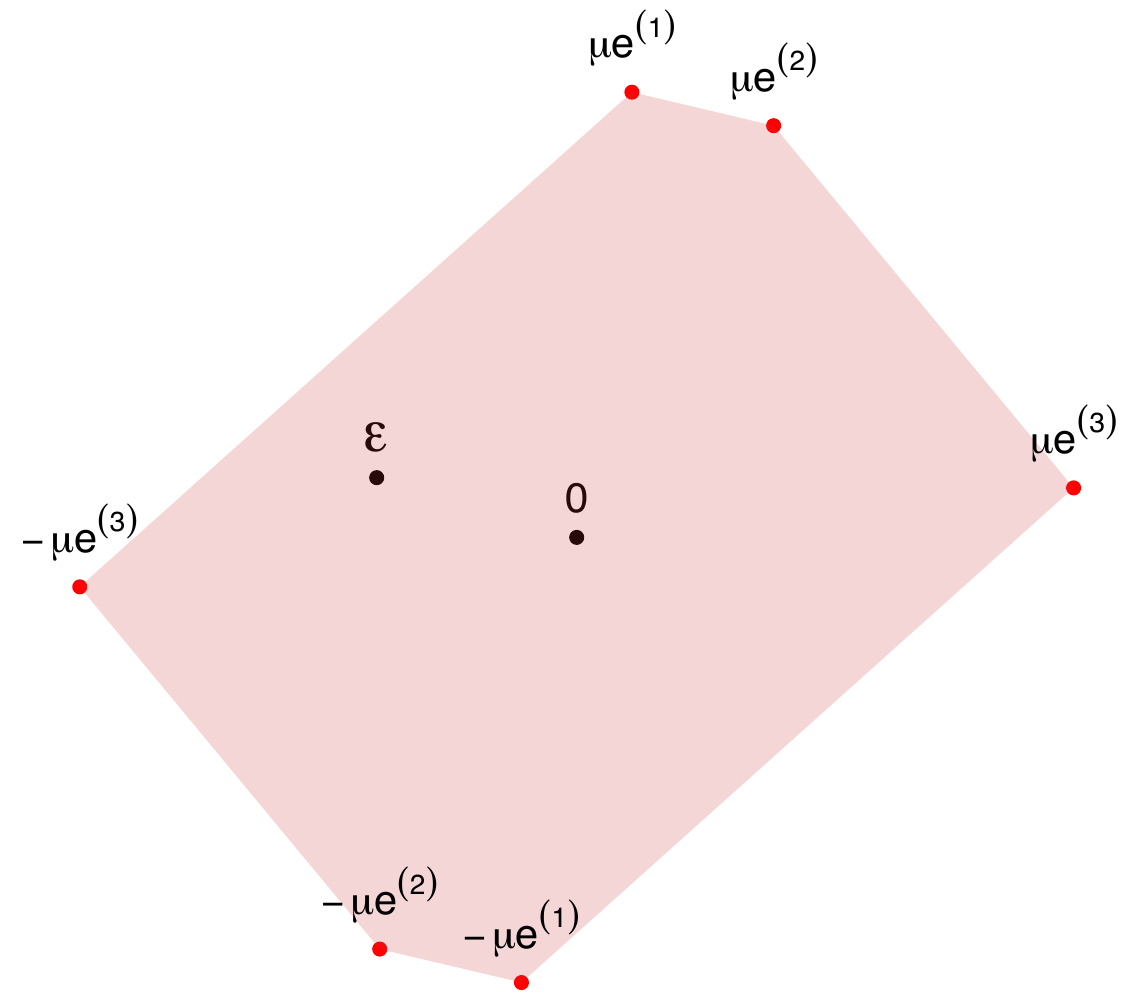}
\caption{  \label{fig:convex} Left:  the convex region~(\ref{def:Nm}) around the observations $\mb Y$ that contains $\mathbb{E}(\mb Y)$ with high probability. Right: the equivalent property~(\ref{def:conhull}) that the noise and the (random) convex region $N_{m,\mu}$ have to satisfy. Here the number of vertices is $m=3$ (the negative counterparts are not counted in $m$).   }
\end{figure}
Define the convex hull of these $2m$ vectors, if rescaled by a factor $\mu\ge 0$ by $N_{m,\mu}  :=   \mbox{convex hull}( \mu \mb E) $, 
where the convex hull is understood column-wise, such that it can be parameterised as \begin{equation}\label{def:Nm} N_{m,\mu} \;=\; \Big\{ \mbb \eta \in\mathbb{R}^n \Big|   \mbb \eta= \mu\cdot \mb E \mbb \gamma \mbox{ for some } \mbb \gamma\in \mathbb{R}_+^{2m} \mbox{ with }\sum_{k=1}^{2m} \mbb \gamma_k\le 1 \Big\}.\end{equation}
The origin is by construction an element of $N_{m,\mu}$.
An illustration of $N_{m,\mu}$ for $m=3$ vertices in a two-dimensional problem is given in the right panel of Figure~\ref{fig:convex}.
Suppose the number of points $m$ and the scaling factor $\mu$ are chosen such that 
\begin{equation}\label{def:conhull} \mathbb{P}( \mbb \varepsilon\in N_{m,\mu} ) \ge 1-\alpha .\end{equation}
If the noise distribution is completely known, then one can for example simulate from the \lhs in~(\ref{def:conhull}) to ensure that the constraint is satisfied. For a known noise distribution, the noise vector is clearly a pivotal quantity and this is exploited in the argument above and could possibly be extended to fiducial-type inference \citep{cisewski2012generalized,wang2012fiducial,taraldsen2013fiducial}.
We will return later to the question of the choice of $m$ and $\mu$ if the variance of the noise is unknown (as it will be in practice).
If we chose $N_\alpha$ in~(\ref{def:conhull3}) as $N_{m,\mu}$, defined in~(\ref{def:Nm}), with appropriate values of $\mu$ and $m$, the region $C_\alpha$ in~(\ref{def:C}) becomes
 \begin{align}\label{def:C2} C_{m,\mu} \; :=\; \Big\{  (\mbb \beta,\mbb \gamma) \in (\mathbb{R}^{p} ,\mathbb{R}^{2m}_+ ) \Big|\; & \sum_{k=1}^{2m} \mbb \gamma_k =1    \mbox{ and }  \mbb \beta = b(\mb X,\mb Y +\mu\cdot \mb E\mbb \gamma )     \Big\}.\end{align}

As discussed above, the set $C_{m,\mu}$ is not necessarily convex. To obtain a convex relaxation, let \begin{equation} \label{eq:bpk} \mbb \beta^{(k)} = b(\mb X,\mb Y + \mu\cdot \mb E \mbb \gamma^{(k)})\end{equation} be the Basis Pursuit solution for a vector $\mb Y+\mu \cdot \mb E \mbb \gamma^{(k)}$, where $\mbb \gamma^{(k)}_j= 1\{j=k\}$ for $1\le j,k\le 2m$. In Figure~\ref{fig:convex}, this would correspond to the Basis Pursuit solution at the six vertices of the shaded regions in the left panel.  Let $l_k= \|\mbb \beta^{(k)}\|_1$ be the $\ell_1$-norm of the corresponding Basis Pursuit solutions for $k=1,\ldots,2m$.
By definition of Basis Pursuit, we have, as long as $\min_{1\le k\le m} \mbb \gamma_k \ge 0$ and $\sum_k^{2m}\mbb \gamma_k\le 1$, 
 \[\| b(\mb X,\mb Y+ \mu \cdot \mb E \mbb \gamma ) \|_1 \; \le \; \sum_{k=1}^{2m} \mbb \gamma_k l_k  ,\]
since the convex mixture $\tilde{\mbb \beta} =\sum_{k=1}^{2m} \mbb\gamma_k \mbb\beta^{(k)}$ is a feasible solution to $\mb X\tilde{\mbb \beta}=\mb Y +\mu\cdot \mb E \mbb\gamma$, with $\ell_1$-norm bounded by the convex mixture of the individual $\ell_1$-norms, $\|\tilde{\mbb \beta}\|_1\le \sum_{k=1}^{2m} \mbb \gamma_k \|\mbb \beta^{(k)}\|_1$.
Using this bound, we define a convex relaxation of~(\ref{def:C2}) as 
 \begin{align}\label{def:barC} \bar  C_{m,\mu} \; :=\; \Big\{  (\mbb \beta,\mbb \gamma) \in (\mathbb{R}^{p} ,\mathbb{R}^{2m}_+ ) \Big|\; & \sum_{k=1}^{2m} \mbb \gamma_k =1   \mbox{ and }  \|\mbb \beta\|_1 \le \sum_{k=1}^{2m} \mbb \gamma_k l_k  \mbox{ and } \mb X\mbb \beta = \mb Y+\mu\cdot \mb E\mbb \gamma    \Big\}.\end{align}
The optimisation in~(\ref{def:stat}) with the set $C_\alpha$  can then be cast as a linear programming problem. The general case of $q\ge 1$ can be solved with standard convex optimisation routines but we will not go into more detail and mostly discuss the case of $q=1$. If we are just interested in testing the null hypothesis $\mbb \beta_G\equiv \mbb 0$ rather than building confidence intervals, we can use either value of $q\ge 1$  and the choice $q=1$ offers the computationally most efficient solution for testing. 

In the following, we will always understand the estimator $T_G$ to be the solution of~(\ref{def:stat}), using the set $\bar C_{m,\mu}$ in~(\ref{def:barC}). We will discuss in the following how the values of $m$ and $\mu$ can be chosen to guarantee~(\ref{def:conhull}) when the noise level of the error distribution is unknown.

\subsection{Unknown noise level}
So far, we have assumed that we know the noise distribution and can thus guarantee~(\ref{def:conhull}) to be true.  Even if the distributional form is approximately known, the noise level itself is in general unknown in practice.
A challenge when implementing the procedure is thus that we have to 
determine the number of vertices $m\in\mathbb{N}$ and the scale factor $\mu\ge 0$ in a way such that~(\ref{def:conhull}) is satisfied at the desired level $\alpha$. If either the underlying distribution of the noise were known or the $\ell_2$-norm  of the realised noise were known, it is straightforward to satisfy constraint~(\ref{def:conhull}).
\begin{itemize}
\item[(i)] If the distribution of the noise $\mbb \varepsilon$ were known, then a suitable strategy uses a fixed scaling factor \begin{equation} \label{eq:muC} \mu \;=\;  C q_{1-\alpha} (\|\mbb \varepsilon\|_2),\end{equation} where $C>1$ is a fixed constant and $q_{1-\alpha} (\|\mbb \varepsilon\|_2) $ the ($1-\alpha$)-quantile of the distribution of  $\|\mbb \varepsilon\|_2$. We could then determine the \lhs of~(\ref{def:conhull}) as a function of the number of vertices $m$ by simulation and choose $m$ large enough to satisfy~(\ref{def:conhull}). If $C>1$, the number of vertices necessary will always be finite and we will use a default value of $C=3$.
\item[(ii)] If the $\ell_2$-norm $\|\mbb \varepsilon\|_2$  of the realised noise were known, one could choose as scaling factor a small multiple, $\mu = C\|\mbb \varepsilon\|_2$ with $C>1$  with a default again of $C=3$) and  choose $m$ so that~(\ref{def:conhull}) is satisfied. There will always be a finite number of $m$ for which the property is satisfied as long as $C>1$ as the convex hull will then contain the $\ell_2$-ball with radius $\|\mbb \varepsilon\|_2$ for $m\rightarrow \infty$).
\end{itemize}
In general, neither the exact distribution nor the realised norm $\| \mbb \varepsilon\|_2$ are  known.

 Assume that an initial estimator of $\hat{\mbb \beta}$ is available which has not made use of the current data (we return to an implementation using sample splitting further below). The residuals are then  $\mb R= \mb Y  - \mb X\hat{\mbb \beta}$. The $\ell_2$-norm of the residuals is $\| \mb R\|_2 = \| \mb d +\mbb \varepsilon\|_2$, where $\mb d:=\mb X (\mbb \beta^* - \hat{\mbb \beta} )$. The norm of the residuals $\|\mb R\|_2$ often provides a good upper bound for $\|\mbb \varepsilon\|_2$, although it can obviously happen that $\|\mb R\|_2 < \|\mbb \varepsilon\|_2$ and we will have to work a bit more to deal with this scenario. However, since $\|\mb R\|_2$ is typically approximately equal and often slightly larger than $\|\mbb \varepsilon\|_2$, we will  fix the scaling factor $\mu$ at $3 \|\mb R\|_2$ for the following and then  have to determine the number of vertices $m$ such that 
\begin{equation}\label{def:conhull2} \pi(\mb d) := \mathbb{P}\big( \mbb \varepsilon\in N_{m,3\|\mb d+\mbb \varepsilon\|_2} \big) \ge 1-\alpha. \end{equation}
The issue is that the vector $\mb d=\mb X(\mbb \beta^*-\hat{\mbb \beta})$ is unknown. In can often be assumed to be small, but even this is hard to establish with tight bounds in practice. We can, however, use the rotational invariance of the noise distribution to see that $\pi(\mb d)$ in~(\ref{def:conhull2}) is just a function of the size $\kappa:=\|\mb d\|_2$ of $\mb d$ and not its orientation $\mb d/\|\mb d\|_2$, and hence $\pi(\mb d) = \pi( \kappa \mb u)$, where $\mb u$ is a vector with unit length and can without limitation of generality be chosen to be the $n$-dimensional vector with entries $\mb u_i=1\{i=1\}$ for all $1\le i\le n$. The \lhs of~(\ref{def:conhull2}) can be bounded by
\begin{align}\label{def:conhull3} \min_{\mb d \in \mathbb{R}^n} \pi(\mb d)\; \ge \; \min_{ \kappa\ge 0} \pi(\kappa \mb u ) & =\;  \min_{ \kappa\ge 0}  \mathbb{P}\big( \mbb \varepsilon\in N_{m,3  \|\kappa \mb u+\mbb \varepsilon\|_2} \big) \nonumber \\ &\ge\;  \mathbb{P}\big( \mbb \varepsilon\in N_{m,3 \min_{\kappa\ge 0} \|\kappa \mb u+\mbb \varepsilon\|_2} \big) , \end{align}
where the inequality holds 
since the origin is contained in all $N_{m,\mu}$ by construction and we thus have that $N_{m,\mu_1}\subseteq N_{m,\mu_2}$ for all $\mu_1\le \mu_2$. It remains to show that  the \rhs in~(\ref{def:conhull3}) is greater than $1-\alpha$. Let $\mu_*\ge 0$ be defined as 
\[ \mu_*^2 = \left\{ \begin{array}{cl}   \|\mbb \varepsilon\|^2_2 & \mbox{ if } \mbb \varepsilon_1\ge 0 \\ \|\mbb \varepsilon\|^2_2 - \mbb \varepsilon_1^2  & \mbox{ if } \mbb \varepsilon_1 < 0     \end{array}   \right. \]
By definition,  $\min_{\kappa\ge 0} \|\kappa \mb u+\mbb \varepsilon\|_2 = \mu_*^2$.
If we now choose $m$ so that 
\begin{equation}\label{def:conhull4}  \mathbb{P}( \mbb \varepsilon\in N_{m,3 \mu_*} ) \; \ge\;  1-\alpha, \end{equation}
then we guarantee~(\ref{def:conhull2}) and thus~(\ref{def:conhull}). Crucially, the \lhs of~(\ref{def:conhull4}) can be determined by simulation, where both the noise $\mbb \varepsilon$ and the convex region $N$ are randomly generated. Specifically, the unknown bias $\mb d$ of the initial estimator does not enter into~(\ref{def:conhull4}).

 \begin{table}
\begin{center}
\caption{ \label{tab:1}The number of vertices per sample point, $m/n$, necessary  to achieve the desired confidence~$1-\alpha$ for sample size~$n$ for unknown noise level.    }
\begin{tabular}{l cccc cccc} \hline
$n=$&  $5$  & $10$  & $15$&$20$& $25$  & $30$  & $40$&$50$ \\ \hline
$\alpha= .05$ & 2.8&3.4 & 4.7 & 6.5 & 8.8 & 12 & 23.5 & 41.8 \\
$\alpha= .025$ & 3.4 & 3.9 & 5.2 & 7.1 & 9.7&13.2 & 25.8 & 46 \\
$\alpha= .01$ & 5.6& 4.8 & 6 & 8.6 & 10.7&14.5 & 28.4 & 50.6 \\
$\alpha= .005$ & 14.6&5.5 & 7 & 9.5 & 11.8 & 16 & 31.2 & 55.7 
\end{tabular}
\end{center}
 \end{table}

 Up to this point, we have not made use of the distributional form of the error distribution except for rotational invariance.
If we now assume a Gaussian distribution with unknown noise level~$\sigma^2$, then we can use simulations of the \lhs of~(\ref{def:conhull4}) to determine the number of vertices~$m$ necessary as a function of sample size $n$ only. Note that the \lhs is invariant under a change in~$\sigma^2$ (since the region~$N$ scales linearly with the noise level) and we can simulate under, say,~$\sigma=1$ or any other arbitrary noise level.

Some results are given in Table~\ref{tab:1}. We use as scaling factor a constant~$\mu=3\|\mb R\|_2$, that is three times the $\ell_2$-norm of the residuals. The number of vertices per sample,~$m/n$, to reach a guaranteed level~$\alpha$ in ({\ref{def:conhull}) were computed, using 5000 simulations. The number of vertices~$m=m(n)$ necessary for a given sample size is generally increasing super-linearly in~$n$, manifesting itself in a monotonous increase of the ratio $m(n)/n$ in Table~\ref{tab:1}. The only exception are very small values of~$n$ and small values~$\alpha$, where the ratio~$m(n)/n$ is decreasing up to a sample size~$n=10$ and increasing afterwards (since the difference between~$\mu_*$ and~$\|\mbb\varepsilon\|_2$ can be substantial for a small sample size).

We reiterate that  the values in Table~\ref{tab:1} are valid  for all noise levels under the assumption of Gaussian noise, irrespective of how the initial estimator~$\hat{\mbb \beta}$ was computed under which the norm of the residuals~$\mb R$ are derived (as long as the estimator did not make use of the current data). If the initial estimator~$\hat{\mbb \beta}$ is very imprecise, the coverage will in general be better than $1-\alpha$ (since~$\|\mb R\|_2$ will be substantially larger than~$\|\mbb \varepsilon\|_2$). The procedure will thus be unduly conservative if the initial estimator has a substantial error, but the level is guaranteed in all circumstances.

\subsection{Summary of the procedure}\label{section:algorithm}
In summary, the procedure works as follows, given an initial estimator $\hat{\mbb \beta}$ that has been computed on a separate dataset.
\begin{enumerate}
\item Compute the residuals $\mb R=\mb Y- \mb X\hat{\mbb \beta}$  and set $\mu=3\|\mb R\|_2$.
\item Set the number of vertices $m=m(n)$ to satisfy~(\ref{def:conhull4}), for example using the values in Table~\ref{tab:1}.
\item Simulate the $2m$ vertices as in~(\ref{def:M}) to get the $n\times (2m)$-dimensional matrix $\mb E$.
\item Solve estimator~(\ref{def:stat}) over the convex area~(\ref{def:barC}) with the values of $\mu$ and $m$ as found in steps~1 and~2. If $T_G=0$ in~(\ref{def:stat}), we cannot reject the null hypothesis $\mbb \beta_G\equiv\mbb 0$. Otherwise  $[T_G,\infty)$ is a non-trivial one-sided $1-\alpha$ confidence interval for $\|\mbb \beta_G\|_1$.
\end{enumerate}

Given an initial estimator $\hat{\mbb \beta}$, the procedure can thus give a $1-\alpha$ one-sided confidence interval for the $\ell_1$-norm of the coefficients in the group  under without making any assumption on the design matrix. 

We will address possible generalisations of (b) further below, after discussing an integrated procedure that computes the initial estimator and the confidence intervals on the same dataset by using repeated data-splitting, and projections for faster computation. 
The final procedure is implemented as function {\tt groupLowerBound} in the {\tt R}-package {\tt hdi} \citep{R}.

\subsection{ Data splitting}
The procedure as above depends on an initial estimator that is not making use of the available data. In practice, we want to derive the estimator on the same dataset. One can use data splitting to derive the confidence interval in the spirit of \citet{wasserman2009high} and \citet{meinshausen09pvalues}. For a given split of the $n$ samples into two parts of equal size (or as close as possible if $n$ odd), we compute the initial estimator $\hat{\mbb \beta}$ on the first part of the data and use it on the second half according to Section \ref{section:algorithm}.

The randomness introduced by this data split is unnecessary. And so is the randomness introduced by the selection of the random support vectors $\mbb e^{(1)}, \ldots, \mbb e^{(m)}$ in the construction of $\mb E$ in~(\ref{def:M}). 
We can repeat the data splitting $K$ times to obtain the statistics $T^{(1)}_G, \ldots, T^{(K)}_G$ according to~(\ref{def:stat}) and then obtain the $1-\alpha$ confidence 
for $\|\mbb \beta_G\|_1$ as 
\begin{equation}\label{median} [ (1-\epsilon)\mbox{-quantile}_{1\le k\le K} T^{(k)}_G,\infty),\end{equation}
which retains the desired $1-\alpha$ coverage  if individual tests in the $K$ splits are conducted at level $q \epsilon $, as shown in \citet{meinshausen09pvalues}. For example, we can use $\epsilon=0.5$ and use the median of all realisations of $T^{(k)}_G$. Empirically, it tends to be more powerful (yet also computationally more demanding) to use higher quantiles of the distribution. We settle here for a  compromise of $\epsilon=0.1$ and use  the $90\%$-quantile, where the individual test are conducted at level $\alpha/10$. The aggregated result will then not depend on an arbitrary split of the dataset and the random sampling of the support vectors.

\subsection{Heavy-tailed error distributions}
Table~\ref{tab:1} shows the necessary number of vertices $m=m(n)$ to guarantee the coverage property~(\ref{def:conhull}) and was derived under the assumption of Gaussian noise with unknown noise level $\sigma^2$. For other error distributions, we have to consider two cases. As long as the error is still rotationally invariant, we can simply simulate in the same way as above for a Gaussian error in~(\ref{def:conhull4}) to obtain a suitable value of $m=m(n)$ as we have only made use of rotational invariance leading up to~(\ref{def:conhull4}). For non-rationally invariant distributions, we will have to find other ways of taking the step from~(\ref{def:conhull2}) to~(\ref{def:conhull3}) that deals with the unknown bias $\mb d$ in the prediction of the initial estimator. It seems conceivable such steps can be found for a variety of other distributions.
One possibility is to find a probabilistic lower bound $\hat{\mu}$ for $\|\mbb \varepsilon\|_2$ such that $\mathbb{P}(\hat{\mu} > \|\mbb \varepsilon\|_2) <\alpha/2$. We would then use the ball $N_{m,3\hat{\mu}}$ and would have to choose the number of vertices such that~(\ref{def:conhull4}) holds with probability at least $1-\alpha/2$, where we can use $\mu_*= 3\|\mbb \varepsilon\|_2$ in~(\ref{def:conhull4}), which is known in the simulation. Combining the two possible errors with a union bound will guarantee overall the desired level. Another possibility is to use support vectors along the axes instead of randomly sampled support vectors on the sphere. Then the convex region $N_{m,\mu}$ will correspond exactly to the region with bounded $\ell_1$-norm, which might be more suitable for heavy tailed error distributions and yield less conservative estimators. 

\subsection{Projection}
Property~(\ref{res:alpha}) and Theorem~\ref{th:1} rests solely on the fact that $\mbb \beta^*$ is by assumption the Basis Pursuit solution to the noiseless signal, that is 
\[ \mbb \beta^* \; =\; b(\mb X, \mathbb{E}(\mb Y)) ,\] with the Basis Pursuit solution defined as in~(\ref{def:b}).
Let $\mb A$ be any linear operator $\mb A \in\mathbb{R}^{s \times n}$ with $s\le n$. Let  $  \mbb \beta^{*,\mb A}$ be the sparsest solution of the projected data
 \begin{equation}\label{BP} \mbb \beta^{*,\mb A} \; =\; b(\mb A\mb X, \mb A \mathbb{E}(\mb Y)). \end{equation}
If $\mbb \beta^* \equiv \mbb \beta^*_{\mb A}$ (we will discuss conditions for this further below),
property~(\ref{res:alpha}) and Theorem~\ref{th:1} are still valid. 
Moreover, if $\mbb \varepsilon$ has a rotationally invariant distribution, and we use a matrix~$\mb A$ with orthogonal and unit-length rows, then the error $\mb A\mbb \varepsilon$ will again be rotationally invariant and independent. Specifically, if $\mbb \varepsilon_i$, $1\le i\le n$ are \iid Gaussian, then $(\mb A\mbb \varepsilon)_j$, $1\le j\le s$  will again be \iid Gaussian (it is not strictly necessary that the columns of $\mb A$ have unit-norm but we will assume so anyway for simplicity in the following). 

Working with the projected data has two potential advantages
\begin{itemize}
\item[1)] Computing the estimator~(\ref{def:C}) is  faster since the problem is now only $p+2m(s)$-dimensional instead of the original $p+2m(n)$-dimensional problem, where $m(n)$ is the number of vertices necessary to guarantee~(\ref{def:conhull}). Since $m(\cdot)$ generally scales super-linear in its argument (see Table~\ref{tab:1}), this can lead to considerable computational advantages.
\item[2)] The convex relaxation in~(\ref{def:barC}) might be less conservative in the lower-dimensional setting and the procedure hence more powerful.
\end{itemize}
A potential issue is the loss of power if the projection $\mb A$ is not chosen suitably. For example,~$\mb A$ can correspond to subsampling~$s$ observations out of the total of~$n$ observations if each row of~$\mb A$ is identically 0 except for a single 1 entry and this choice of the projection will lead to a substantial loss in power if~$s\ll n$.
This problem is alleviated, however, if we chose the estimated signal direction $\mb X\hat{\mbb \beta}$ as one of the rows of $\mb A$. If we are using a projection in the numerical results, we will hence assume that the rows of $\mb A$ are the unit-norm base vectors of the space spanned by $\mb X\hat{\mbb \beta}$ and~$s-1$ vectors in~$\mathbb{R}^n$, whose entries are drawn \iid from a standard Gaussian distribution. 

 The numerical results suggest that the procedure is very insensitive to the choice of $s$ in general.
We will give exact conditions necessary for success in the following section.

\section{Estimation accuracy}
We will look at properties of the design that need to be satisfied for the estimator~(\ref{def:stat}) to have power close to 1 to detect groups of variables that contribute substantially to the overall signal. We will use the projected data approach, as discussed in the last section, with an orthonormal projection matrix. While we do not need to make assumptions on the design to show the correct coverage of the confidence interval $[T_G,\infty)$ as in Theorem~\ref{th:1}, some additional assumptions on the design are needed for showing the estimation accuracy. For example, if there exists a variable outside of  $G$ that is perfectly correlated with a variable in group $G$, we can not hope to get sharp bounds on $\|\mbb\beta^*_G\|_1$ as the problem is not identifiable. The same situation does not pose a problem for coverage, though, as the method would always choose the most conservative possibility among all possible solutions. The design assumption we will need to impose are, however, weaker than all known conditions to detect individual variables in the high-dimensional setting.

\subsection{Compatibility condition} The weakest condition for rates of convergence and variable-wise confidence intervals rest on the \emph{compatibility condition} \citep{van2009conditions}. We  will be able to weaken the condition for the group case. Assume $S_0$ to be the set of variables that have a non-zero effect $S_0=\{k: \beta^*_k\neq 0\}$. (Alternatively, we could let $S_0$ be the set of variables that have a sufficiently large non-zero effect and which we do want to detect with the test.) Let $L>0$ be a constant.  The \emph{compatibility constant} $\phi_{cc}$ is defined as in \citet{van2009conditions} for a design $\mb X$  as
\begin{equation}\label{eq:cc} \phi_{cc}^2(L) := \; \min\big\{ |S_0| \|\mb X \mbb \beta \|_2^2 :\;\; \|\mbb \beta_{S_0^c}\|_1 \le L \|\mbb \beta_{S_0}\|_1 \mbox{  and  } \|\mbb\beta_{S_0} \|_1 \ge 1 \big\} .\end{equation}
The multiplication with $|S_0|$ could be left out but facilitates comparisons with eigenvalues constrained by the $\ell_2$-norm instead of the $\ell_1$-norm.
The compatibility condition requires $\phi_{cc}$ to be bounded away from zero for, typically,  a value $L\ge 3$.
All known conditions for consistency of the Lasso and convergence of confidence intervals imply the \emph{compatibility condition} for some value of $L\ge 1$  \citep{van2009conditions}.

\subsection{Group effect compatibility condition} \label{section:gcc}
The \emph{compatibility condition} is really geared towards detection of individual variables. It often fails for real data due to high correlation between variables. Here, we define the \emph{group effect compatibility constant} that leads to weaker assumptions if we are just interested in the effect of a group of variables. 
\begin{definition}[Group effect compatibility constant]\label{def:gcc}
The group effect compatibility constant $\phi_{gcc}$ for a  group $G\subseteq\{1,\ldots,p\}$ is defined as 
\begin{equation}\label{eq:scc}  \phi_{gcc}^2(L, G) := \; \min\big\{ |S_0| \|\mb X \mbb \beta \|_2^2 :\;\; \|\mbb \beta_{G^c \cap S_0^c}\|_1 \le L (\|\mbb \beta_{S_0}\|_1-\|\mbb \beta_{G\cap S_0^c}\|_1 ) \mbox{  and  } \nu_{ G}(\mbb\beta) \ge 1 \big\} ,\end{equation}
where  \begin{equation}\label{eq:nu} \nu_{G}(\mbb\beta) :=  \min_{\varsigma \in \mathcal{S}} \sum_{k\in G} \varsigma_k  \beta_k ,\end{equation} 
and $\mathcal{S}\subseteq[-1,1]^p$ is defined as the subgradient of the $\ell_1$-norm evaluated at $\mbb \beta^*$:
\[ \mathcal{S} \;:=\; \big\{ \varsigma \in [-1,1]^p: \varsigma_k = \mbox{sign}(\beta^*_k) \mbox{ if } k\in S_0 \mbox{ and } \varsigma_k\in [-1,1] \mbox{ otherwise}\big\}.\]
\end{definition}
The dependence on the design and on $\mbox{sign}(\mbb \beta^*)$ has been suppressed for sake of notational simplicity  in the definition of compatibility constant. The constant is not to be confused with the \emph{group Lasso compatibility constant}, which is adapted to study the group Lasso estimator \citep{buhlmann2011statistics} and not directly comparable to the compatibility constants discussed here.

 For all $G\subseteq\{1,\ldots,p\}$ and $L\ge 1$, the \emph{group compatibility constant} is lower-bounded by the \emph{compatibility constant},
\begin{equation}\label{eq:greater} \phi_{gcc}^2(L,G) \ge \phi_{cc}^2(L). \end{equation}
This property  follows from the following observations.  For any  group $G\subseteq\{1,\ldots,p\}$, the bound  $\nu_{G}(\mbb\beta) \le \|\mbb\beta_{S_0}\|_1$ holds true. The condition $\nu_{ G}(\mbb\beta) \ge 1 $ thus implies the corresponding condition $\|\mbb\beta_{S_0} \|_1 \ge 1$ in the \emph{compatibility constant}. Furthermore, for any $L\ge 1$, 
 $\|\mbb \beta_{G^c \cap S_0^c}\|_1 \le L (\|\mbb \beta_{S_0}\|_1-\|\mbb \beta_{G\cap S_0^c}\|_1 )$ implies the inequality  
$ \|\mbb \beta_{S_0^c}\|_1 \le L \|\mbb \beta_{S_0}\|_1 $. The feasible set in~(\ref{eq:scc}) is thus a subset of the feasible set in~(\ref{eq:cc}), which proves property~(\ref{eq:greater}). 

Any lower bound we impose on the \emph{group effect compatibility constant}~(\ref{eq:scc}) will thus be a weaker condition than the same lower bound on the \emph{compatibility constant}~(\ref{eq:cc}). 
To take an extreme example, assume we have two almost perfectly correlated variables in a group $G$, where both variables have either the same sign or at least not opposite signs  in the regression coefficient $\mbb \beta^*$. As the correlation between the two variables approaches 1, the \emph{compatibility constant} will take the value 0. To see this, one can use a $\mbb \beta$ that uses  coefficients of the same magnitude but opposite signs on the two variables. With this $\mbb \beta$, the $\ell_1$-norm is positive and $\mb X \mbb\beta\equiv 0$ for perfectly correlated variables and the \emph{compatibility constant} thus vanishes. In contrast,  $\nu_G(\mbb\beta)$ remains at 0 for the same vector and the \emph{group effect compatibility constant} will retain a positive value even if both variables are perfectly correlated. (If, however, two almost perfectly correlated variables take the opposite sign in the regression coefficient, then both constants will approach 0 with increasing correlation, as the joint effect of the two variables will be difficult to detect, even if we are just interested in the effect of the group as a whole.)

We note that we can also leverage the hierarchical property of the statistic $T_G$ evident from~(\ref{def:stat}), namely that $T_G\ge T_{G'}$ for all $G,G'\subseteq\{1,\ldots,p\}$ with $G'\subseteq G$. With this hierarchical property we could  weaken the assumption of a lower bound on $\phi_{gcc}^2(L,G)$ by instead assuming a lower bound on $\max_{G':G'\subseteq G} \phi_{gcc}^2(L,G')$. We refrain from developing this further, though, for sake of notational brevity.

\subsection{Assumptions}\label{section:assumptions} Here, we will formulate two conditions on the design  for testing the effect of a set of groups $\mathcal{G}\subset \mathcal{P}(\{1,\ldots,p\})$, where each $G\subseteq \{1,\ldots,p\}$.
\begin{enumerate}
\item[(A~I)] There exists $\varphi_1>0$  a subset $S\subseteq\{1,\ldots,p\}$ of variables with $|S|=s$ such that the corresponding predictor matrix has full rank. In other words, let $(\mb A\mb X)_S$ be the matrix formed by the columns of the full design belonging to variables in the set $S$. The minimal singular value of this matrix is bounded from below by  $\varphi_1>0$.
\item[(A~II)] For $ L= 2$, there exists a $\varphi_2 >0$ for design $\mb A\mb X$ such that 
\begin{align*} 
\min_{G\in \mathcal{G}} \phi_{gcc}^2(L,G)  & \;\ge\; \varphi_2^2.\end{align*}
\end{enumerate}

Some discussion of these assumptions: the first one, (A~I), is a very weak condition since it just requires the existence of a single set of $s$ variables that have a full rank in the projected predictor matrix. The stronger assumption is (A~II). 
It  requires a lower bound on the \emph{group effect compatibility constant} for all groups in $\mathcal{G}$, where the signs derive from the optimal regression coefficient. 
Using the \emph{group effect compatibility constant} $\phi_{gcc} $ makes the assumption much weaker, though, than the typically required lower bound on the \emph{compatibility constant} $\phi_{cc} $ itself  that is necessary for confidence bounds for individual variables  \citep{zhang2011confidence,van2013asymptotically,javanmard2013confidence}. Per definition of the \emph{group effect compatibility condition},
\[ \min_{G\in \mathcal{G}} \phi_{gcc}^2(L,G) \;\ge\; \phi_{cc}^2(L).\]
The value of $L$ for most results is chosen as $L=3$. While the exact value does not matter too much, we chose $L=2$ here but any value larger than 1 would yield similar results. 
 Note that $\phi_{cc}^2(1)>0$ is  also sometimes called the \emph{nullspace condition} used to show equivalence of the $\ell_1$ and $\ell_0$-sparsest solutions to the regression problem, see for example \citet{raskutti2010restricted} and references therein. In particular, the \emph{nullspace condition} implies that the $\ell_1$-sparsest solution, as defined as in~(\ref{BP}), is equal to the $\ell_0$-sparsest solution of the noise-free data.

As discussed in the previous section, the \emph{group effect compatibility constant}  will not be unduly diminished by highly correlated variables that appear in the same group. This is also evident from the empirical results in the  section with numerical results.
In the presence of highly correlated variables, assumption (A~II) can thus be significantly weaker then the otherwise necessary lower bound on the \emph{compatibility constant} as we ask for the effect of whole groups of variables instead of the effect of individual variables.

\subsection{Estimation accuracy}
Under the made assumptions, the procedure will be shown to have a near-optimal detection threshold for groups of variables. Specifically, the lower bound $T_G$ for the $\ell_1$-norm of a group $G$ of variables (or indeed a set of groups) is shown to have a non-asymptotic estimation error that scales like $1/\sqrt{n}$ with sample size. 

\begin{theorem}\label{theo:power}
Let $\mathcal{G}\subseteq \mathcal{P}(\{1,\ldots,p\})$ be a set of groups~$G\subseteq\{1,\ldots,p\}$ such that Assumptions (A~I) and (A~II) are satisfied.  Assume the errors~$\varepsilon_i$, $i=1,\ldots,n$ are independent and either have a mean-zero Gaussian distribution with variance~$\sigma^2>0$ or are sub-Gaussian and are dominated in absolute value by such a distribution and $\mu$ is chosen as in~(\ref{eq:muC}). For any chosen~$\gamma\in (0,0.2)$, with probability at least $1-\gamma$, the lower bound $T_G$ at level $\alpha$ satisfies 
\[ \forall G\in \mathcal{G}:\qquad  T_G \; \ge\; \|\mbb \beta^*_G\|_1 - \frac{ M \sigma }{\sqrt{n}},\qquad\mbox{where   } M^2=    20 s\log\big( \frac{1}{\min\{\alpha ,\gamma\}}\big) \max\big\{ \frac{s}{\varphi_1^2} ,\frac{|S_0|}{\varphi_2^2}   \big\} .\]
\end{theorem}
A proof is given in the Appendix. Note that the bound is valid simultaneous for all groups in the set $\mathcal{G}$. The complementary bound ($T_G \le \|\mbb\beta^*\|_1$ with probability at least $1-\alpha$ simultaneously for all groups) is equivalent to the coverage property shown in Theorem~\ref{th:1}.
Regarding the assumptions:
\begin{itemize}
\item[(a)] The two assumptions (A~I) and (A~II) are just necessary to show the power of the approach. The coverage property of the confidence intervals (Theorem~\ref{th:1}) are still valid even if the two assumptions are not satisfied. 
\item[(b)] The condition about the \emph{group effect compatibility condition} is weaker than the corresponding condition about the \emph{compatibility condition} that is necessary to detect individual variables. 
\end{itemize}

The theorem implies that the power to detect groups will have optimal rates under conditions that can be substantially weaker than the conditions needed for a good power of detecting individual variables. 
The number of variables enters only through the compatibility constant  $\varphi_2$.  The theorem also shows the simultaneous nature of the bound: with a high probability, \emph{all} groups with sufficiently large signal strength will be detected.

\section{Numerical Results}\label{section:numerical}
The procedure is evaluated on simulated and real data.  Sample splitting with 11 splits and 90\%-quantile aggregation as per~(\ref{median}) is used, as implemented in the {\tt R}-package {\tt hdi} and function {\tt groupLowerBound},  where the initial estimator is computed with the 10-fold cross-validated Lasso solution as found in the {\tt glmnet} package \citep{friedman2009glmnet}. The optimisation~(\ref{def:stat}) over the set~(\ref{def:barC}) is implemented with the {\tt limSolve} package \citep{soetaert2009limsolve} in {\tt R} \citep{R}. 

As we are not making any assumption on the design and the examples are high-dimensional in the sense that $p>n$, it could be suspected that the power of the method will be very weak against any reasonable alternative. While there is clearly a price to pay for the assumption-free confidence intervals, we will explore to which extent we can get non-trivial bounds.

\subsection{Simulated data}
Six simple simulations settings are used initially  with $p$ predictor variables and sample size $n$. The predictor variables are randomly drawn (independently across observations) from a Gaussian distribution $\mathcal{N}_p(\mb 0,\mbb \Sigma)$, where $\mbb\Sigma$ has a block structure. Each block consists of $B$ variables. All diagonal elements of $\Sigma$ are equal to 1. The within-block correlation is $\rho_{w}$ and the between-block correlation between all variables is $\rho_{b}$.  The response is simulated as $\mb Y=\mb X\mbb \beta^*+\mbb\varepsilon$, where the noise has \iid Gaussian entries with standard deviation $\sigma>0$. 
The optimal regression vector $\mbb\beta^*$ has 0 entries, except for the $B/2$ even variables $\{2,4,6,\ldots,B\}$ within the first block which have entries of value $\tau>0$ ($B$ is always chosen to be an even number).
The settings of $p,n,B,\rho_w,\rho_b,\tau $  vary across the settings as shown in Table~\ref{settings}. The noise level $\sigma$ is varied for each setting between $\sigma=0.001$ and $\sigma=20$ to study the influence of a varying signal-to-noise ratio on the results. 


\begin{table}
\begin{center}
\caption{\label{settings} The simulation settings} 
\begin{tabular}{lrrrrrr} \hline
Variable  & $p$ & $n$ & $B$ & $\rho_w$ & $\rho_b$ & $\tau$   \\ \hline
Setting (i) & 200 & 50 & 10 & 0.99 & 0.00 & 0.5  \\
Setting  (ii) & 200 & 200 & 20 & 0.999 & 0.00 & 1.0 \\
Setting  (iii) & 1000 & 300 & 50 & 0.8 & 0.10 & 2.0  \\
Setting  (iv) & 200 & 100 & 50 & 0.99 & 0.10 & 2.0  \\
Setting  (v) & 300 & 200 & 100 & 0.999 & 0.00 & 1.0  \\ 
Setting  (vi) & 300 & 200 & 100 & 0.995 & 0.50 & 1.0  
\end{tabular}
\end{center}
\end{table}

Figure~\ref{fig:prob} shows the results for 200 simulations of each setting.
The empirical covariance matrices of a realisation of each setting are shown in the leftmost column of Figure~\ref{fig:prob}. 
The remaining columns show the frequency with which the null hypothesis $H_{0,G}$ is rejected for various groups, starting with the singleton $G=\{2\}$ up to $G=\{1,\ldots,p\}$. For the first four groups, the null hypothesis is false, while it is true for the last group, which contains all variables that have a 0 component in the optimal regression vector. 

Results for three competing methods for $\alpha=0.05$ are shown: the proposed group effect estimator (``G''), with a default value of $s=10$ for the projected dimension. The Ridge effect estimator proposed in \citet{buhlmann2012statistical} (``R'') and the Lasso-based test of individual variables of \citet{van2013asymptotically} (``L''). The latter two tests are designed for individual variables and we reject the group null $H_{0,G}$ if we can reject any of the elements of $G$ after a Bonferroni multiplicity adjustment.
\begin{center}
\begin{figure}
\includegraphics[width=0.98\textwidth]{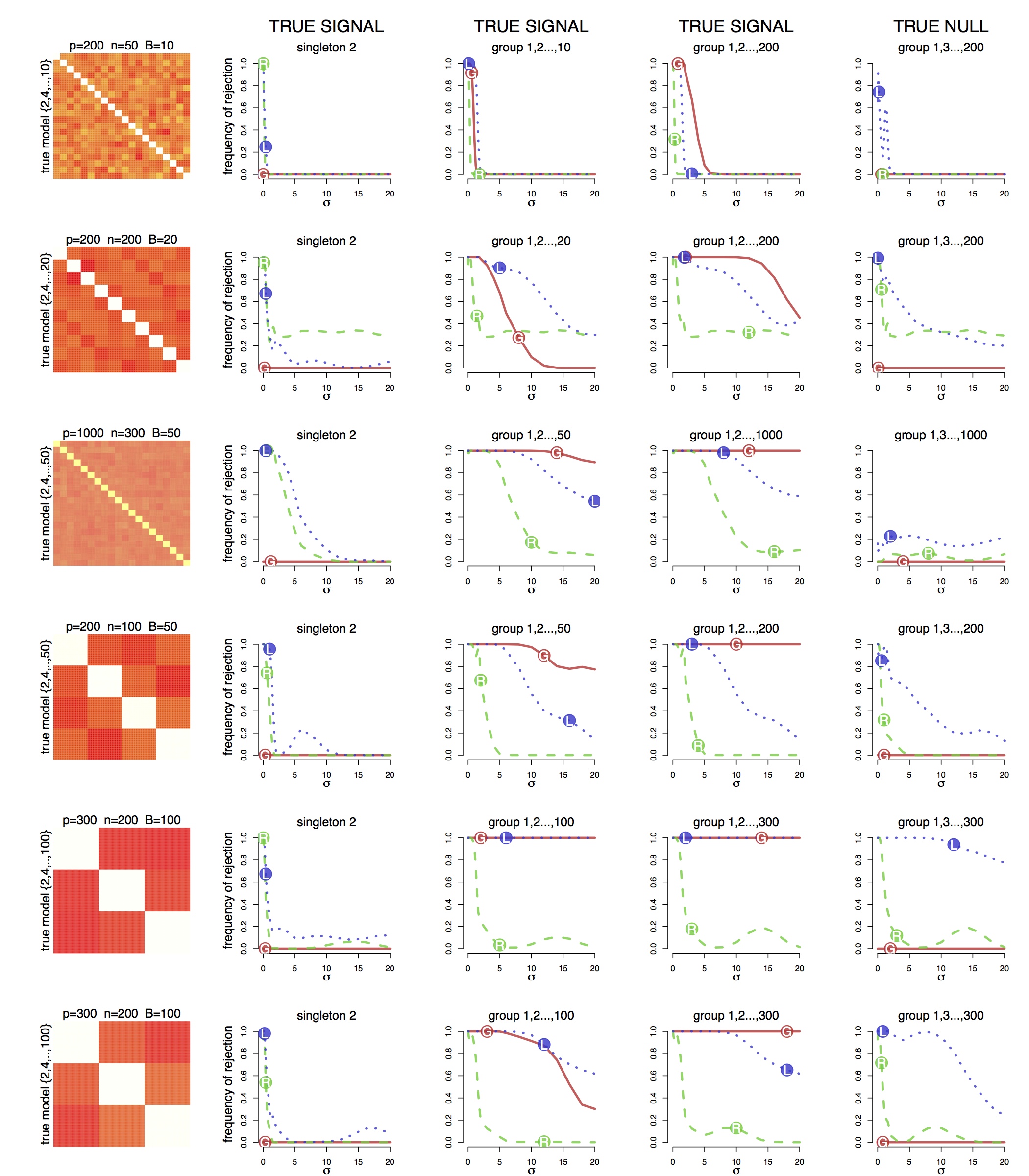}
\caption{ \label{fig:prob}  The six rows correspond to the simulation settings (i)-(vi). The first column shows an empirical correlation matrix, where white corresponds to a value of 1 and orange to 0. The block structure is visible in all settings. 
The remaining columns show the frequency with which the null hypothesis $H_{0,G}$ can be rejected for different groups. The last group contains all variables with vanishing signal and its null hypothesis is true, whereas the null is false for the first three groups.  The results for three different methods are shown:  the proposed group effect estimator (``G''; red solid line), the ridge-based (``R'', green broken line) and lasso-based (``L'', blue dotted line) tests for individual variables that are adapted to the group setting.   }
\end{figure}
\end{center}
The main observations are:
\begin{enumerate}
\item The proposed tests has the correct coverage for all designs (as expected from Theorem~\ref{th:1}) but is conservative: the last group, corresponding to a true null hypothesis, is never rejected. 
\item The other two tests, in contrast, work only under specific design assumptions, which are difficult to verify in practice but which are likely to be violated in these settings due to the high correlation between variables. The type I error (frequency of rejection of the last group, which has a true zero effect) is much higher than the specified $\alpha=0.05$, especially for high signal-to-noise ratios. 
\item The proposed group effect estimator has no power to detect the signal in the individual variable $\{2\}$, whereas the other two tests reject the null hypothesis for this variable for high signal-to-noise ratios (but see the point above: they also frequently reject true null hypotheses).
\item The power to detect signal in groups of variables (the second and third group in Figure~\ref{fig:prob} contains groups with true signal) is often substantially higher with the proposed group effect estimator than with alternatives. This is as expected from Theorem~\ref{theo:power}, as the high correlation between variables in tested groups is compatible with the assumption needed for the Theorem, as discussed in Section~\ref{section:gcc}.
\end{enumerate}


\begin{table}
\begin{center}
\caption{\label{jaccard} Jaccard index of rejections as a function of the projected dimension $s$ compared with the default value  $s=10$} 
\begin{tabular}{lrrrrrrrr} \hline
$s=$  & 2 & 3 & 4 & 5 & 10 & 15 & 20 & 25 \\ \hline
 Setting (i) & 1.00 & 1.00 & 0.96 & 1.00 & 1.00 & 0.96 & 0.95 & 0.96\\
 Setting  (ii) & 1.00 & 1.00 & 1.00 & 1.00 & 1.00 & 1.00 & 1.00 & 1.00\\
 Setting  (iii) & 0.97 & 0.97 & 0.96 & 0.97 & 1.00 & 0.96 & 0.97 & 0.97\\
 Setting  (iv) & 1.00 & 1.00 & 1.00 & 1.00 & 1.00 & 1.00 & 1.00 & 1.00\\
 Setting  (v) & 1.00 & 1.00 & 1.00 & 1.00 & 1.00 & 1.00 & 1.00 & 1.00\\
 Setting  (vi) & 1.00 & 1.00 & 1.00 & 1.00 & 1.00 & 1.00 & 1.00 & 1.00 
\end{tabular}
\end{center}
\end{table}

It remains to study the effect of the dimension $s$ of the projection. The results above in Figure~\ref{fig:prob} were shown for a default value of $s=10$. The results were also computed for $s\in\{2,3,4,5,15,20,25\}$ to study how much they vary across this range. Note that $s=25$ is the maximal possible value since we use sample splitting and the minimal value of $n$ is~50, which means we have then only~25 samples at our disposal within each half of the data.  Instead of re-producing the plots, Table~\ref{jaccard} shows a condensed version. There are~4 examined groups, with~17 noise levels and~200 simulations for each setting. Each setting thus corresponds to~13600 possible rejections. Each values of $s$ corresponds to a subset $R_s\subseteq\{1,\ldots,13600\}$ of rejections made with this value. We record the set of rejections under $s'=10$ and under different values of~$s$ and compare via their Jaccard index as $|R_s\cap R_{s'}|/|R_s\cup R_{s'}|$.  A Jaccard index of 1 thus corresponds to all decisions being identical. For most settings, across all simulations and settings, more then 2000 rejections can be made with the default value and a value of~1 thus corresponds to a remarkable similarity between the results. The lowest value in the table is a Jaccard index of 0.95. Even with this lower value, rejections differ for at most~5\% of all simulation settings and the influence of~$s$ on the properties of the procedure is thus small in practice, justifying a default value that does not have to be adjusted in each new setting. 

The loss or gain in power when using a smaller value of~$s$ is mainly determined by two opposing effects. On the one hand, information is lost by projecting into a lower-dimensional space. This will diminish the power for small values of~$s$. On the other hand, the convex relaxation of the set $C_\alpha$ in~(\ref{def:barC}) becomes less conservative in smaller dimensions. Smaller values of $s$ thus  mitigate the impact of the convex relaxation and can lead to increased power. Which of the two effects is stronger will be problem-specific but the empirical results suggest that the two effect are weak and that the choice of $s$~does not matter much from a statistical perspective. 
Error control is conservative for all values of~$s$: groups that fulfil the null hypothesis are never selected more frequently than in a proportion~$\alpha$ of all simulations.  The speed savings of a lower value of the dimensionality~$s$ of the projection can be considerable, though.  Specifically, computing the estimator for a single realisation and a single data-split takes for $s=5$ in a setting with $p=100$ and $n=50$  an average of 1.07 seconds. Increasing the dimension to the default value of $s=10$ almost triples the computational time to 2.7 seconds, and this increases to 6.46, 13.79 and 29.19 seconds for values $s=15,20$ and $s=25$ (which corresponds to no projection since, with data-splitting, there are just 25 observations in a single half of the data)  on a desktop computer with a single 3.4 GHz CPU. 

It is evident that the procedure provides error control (as already proven in Theorem~\ref{th:1}) and has a decent chance to detect significant groups of variables, even if the variables within a group are highly correlated. In fact, the variables could be perfectly correlated in each block ($\rho=1$) and the results would be almost identical as in setting (i), as expected from Theorem~\ref{theo:power}. This setting of perfect colinearity violates all typical design assumption necessary to get confidence intervals, yet we still have a non-negligible power to detect the effect of the group as a whole with the proposed procedure. 

\begin{center}
\begin{figure}
\includegraphics[width=0.4\textwidth]{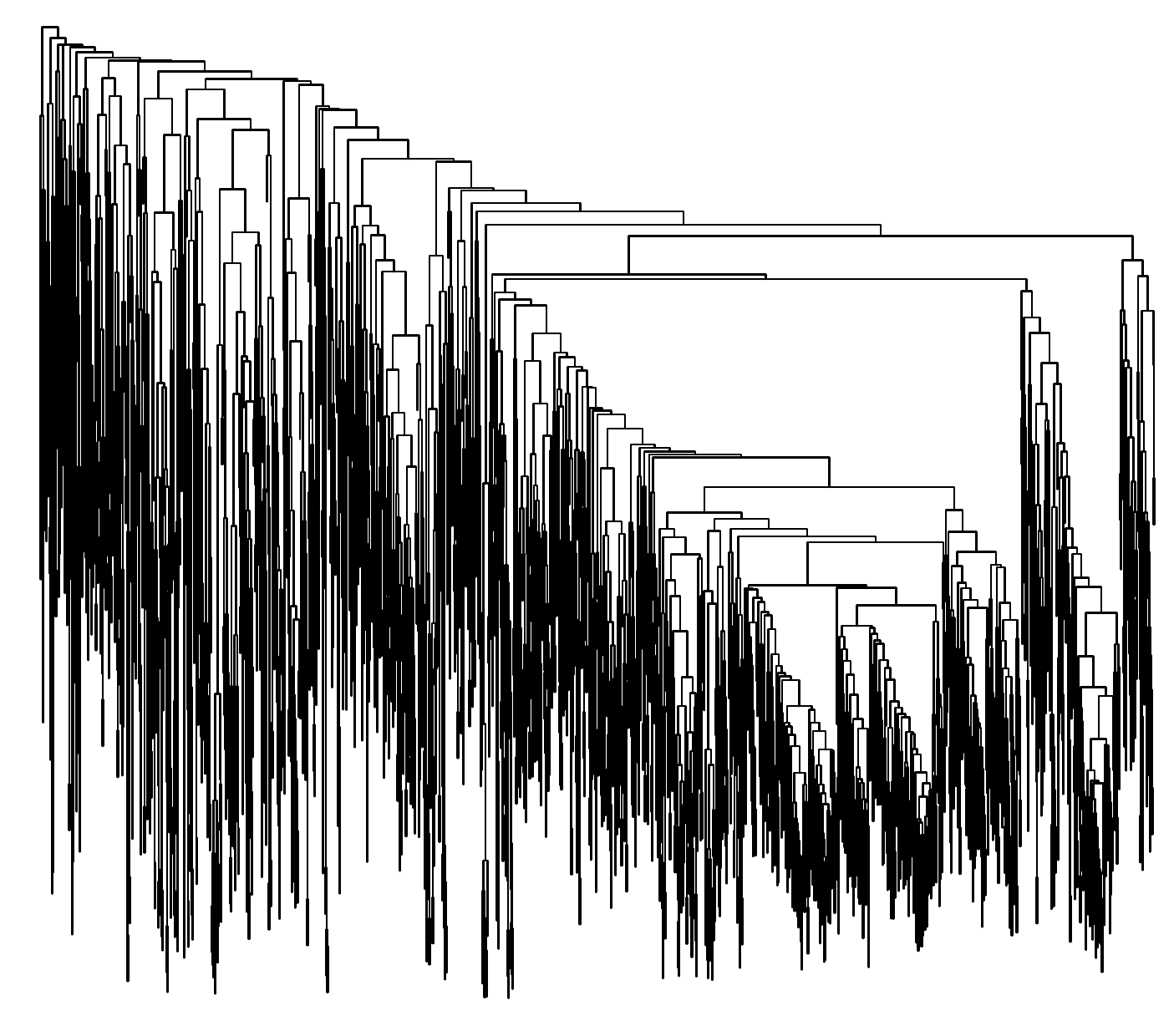}
\includegraphics[width=0.4\textwidth]{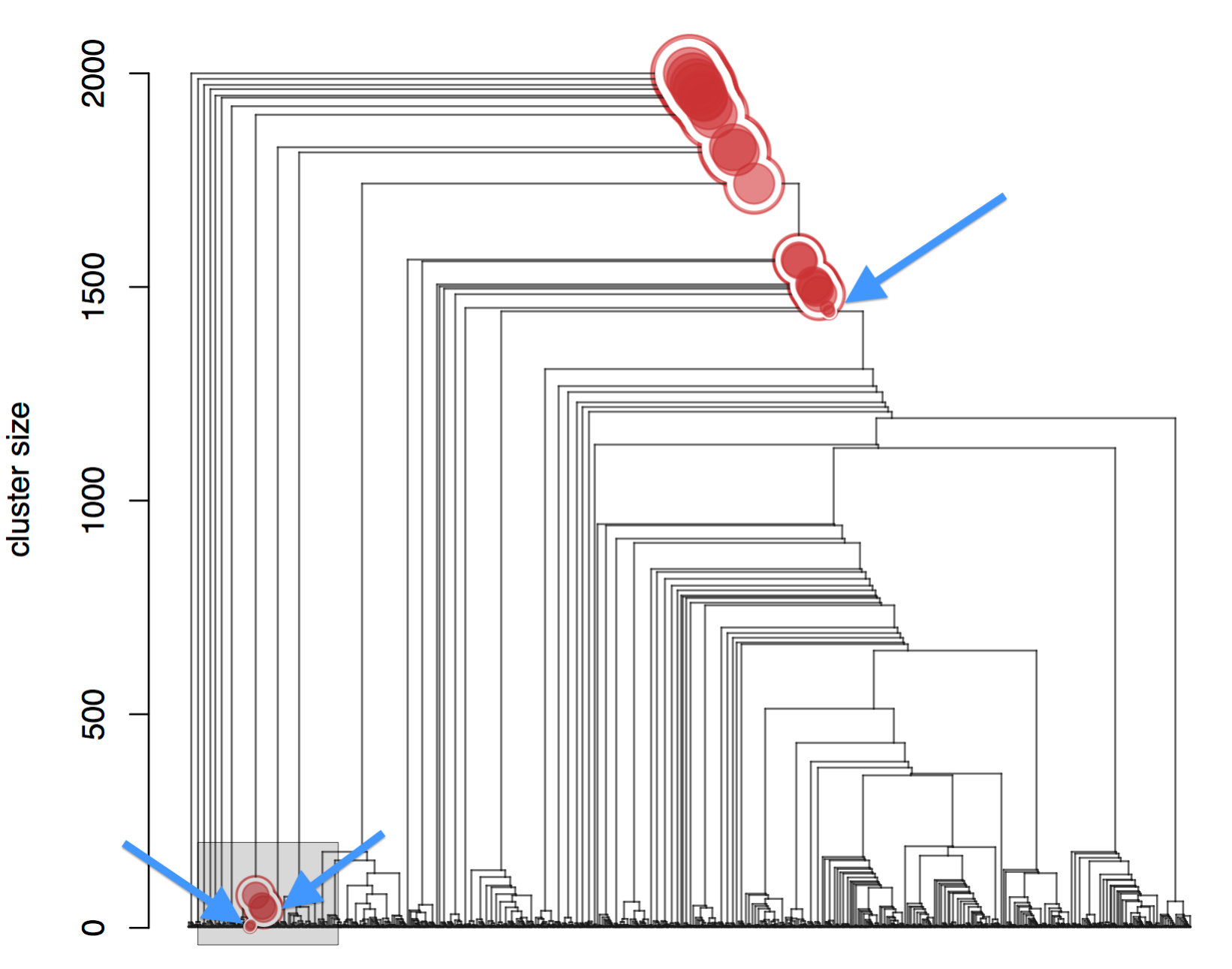}
\includegraphics[width=0.18\textwidth]{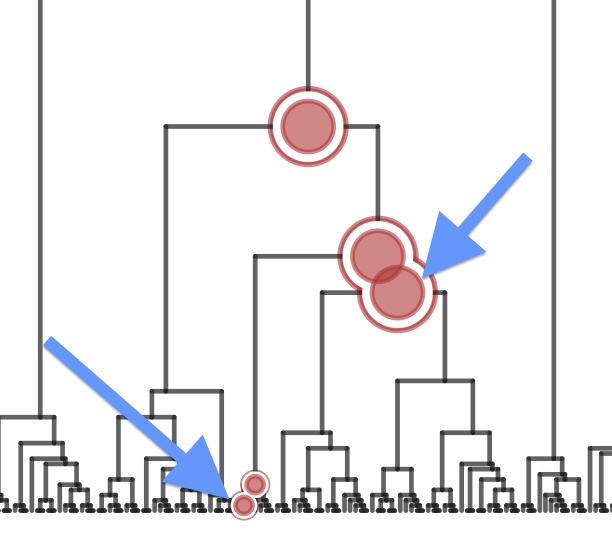}
\caption{ \label{fig:real}  Left: the cluster dendrogram for hierarchical clustering of the 2000 variables with the vitamin expression data (sample size $n=115$). One can test all clusters in a top-down manner. Once a cluster cannot be rejected, all child nodes cannot be rejected as well and the procedure does not need to proceed along the  subtree of a non-rejected cluster.  Middle: the height of each cluster $G$ is shown proportional to the number of its members. The area of the red circles at each cluster node are proportional to the lower bound on the $\ell_1$-norm $\|\mbb\beta^*_G\|_1$ and the area of a cluster node is proportional to the number of variables it contains. Twenty-four clusters have a non-zero bound and three of them are non-overlapping (blue arrows). Right: a close-up of the shaded area in the middle panel, showing that two non-overlapping clusters have been selected in this part of the dendrogram.}
\end{figure}
\end{center}

\subsection{Vitamin expression data}
Next, we take a gene expression dataset, which was kindly provided by DSM Nutritional Products. As described in \citet{meinshausen2008ss}, we have for $n = 115$
samples  a continuous response variable measuring the logarithm of the vitamin B12 production rate of Bacillus Subtilis. Along with this information, the expression levels of $p = 4088$ genes have been measured, covering essentially the whole genome of Bacillus Subtilis. 
The results in \citet{meinshausen2008ss} indicate that Lasso-selection of individual genes is very unstable. We do not touch upon the fact that 
maybe a causal analysis might be more appropriate here by providing more suitable targets for mutation than a regression analysis \citep{maathuis2010predicting}, but simply consider the question whether we can find groups of genes that can be shown to have a significant predictive effect in a sparse linear model. As searching all possible groups of genes will be infeasible, we first cluster the trees with hierarchical clustering, using average linkage. The distance between two genes $i,j$ is defined here, rather arbitrarily, as $1-|\rho_{i,j}|$, where $\rho_{i,j}$ is the empirical correlation between the two genes. 

We can now test all clusters in this tree (including the singletons of individual genes) at level $\alpha$ in an efficient way by a top-down procedure. 
Starting at the root node $G=\{1,\ldots,p\}$, we compute a lower bound for $\|\mbb\beta^*_G\|_1$. If this is non-zero, we can reject the global null hypothesis that there is no predictive power in the optimal linear model. Next, we compute lower bounds for the child cluster nodes of the root node and continue to descend into the tree in this way. Once a cluster is not significant, we can stop searching the whole sub-tree of this cluster node as the lower bound of $\|\mbb\beta^*_{G'}\|_q$ will vanish for all $G'\subseteq G$ if the lower bound of $\|\mbb\beta^*_{G}\|_q$  is zero (irrespective of the value $q\ge 1$). One could also provide a family-wise error control by testing at level $\alpha/p\cdot |G|$ at each cluster $G$ \citep{meinshausen06hierarchical}, which would amount to a Bonferroni-style correction at the level of individual nodes and the usual level $\alpha$ at the root node but we simply test all groups at the same level $\alpha$ here without making a multiplicity adjustment (which would in any case be small with the mentioned scheme in the top layers of the hierarchy). The method is implemented as function {\tt clusterLowerBound} in {\tt R}-package {\tt hdi}.

An example is shown in Figure~\ref{fig:real}, where we have first chosen 2000 of the 4088 genes at random in order to not overwhelm the visual displays.  We also just use three data-splits and a projection to $s=10$ observations to ease the computational burden (it then takes about 2 hours to compute the complete solution on the tree, although it might be  possible to get a more efficient linear programming implementation which could reduce the computational time).  The dendrogram on the right visualises the 24 clusters that have a non-zero lower bound on their $\ell_1$-norms. Three of the clusters are disjoint. The three non-overlapping significant clusters contain 3, 44 (these are the two non-overlapping clusters shown in the close-up on the right) and 1443 genes respectively. 

The root node $G=\{1,\ldots,p\}$ has always the largest lower bound on $\|\mbb\beta^*_G\|_1$, which here has a value of just over 22. It means that the optimal sparse solution has an $\ell_1$-norm of at least 22 at confidence level 0.95. The lower bounds for the $\ell_1$-norms for the three non-overlapping significant clusters of sizes 3, 44 and 1443 have lower bounds on their $\ell_1$-norms of 1.2, 11.4  and 2.1 respectively. The results might look disappointing in that we are not able to reject individual genes. Given that selection of individual genes is very unstable with Lasso estimation \citep{meinshausen2008ss},  it is nevertheless interesting that three non-overlapping clusters $G$ of genes (one just consisting of three genes) can be shown to have a non-zero effect $\|\mbb\beta^*_G\|_1$ without having made any assumption on the design matrix itself.

\section{Discussion}\label{section:discussion}
We have shown that it is possible to construct confidence intervals for the optimal sparse regression coefficients of variables in a high-dimensional setting without making any assumption on the design matrix (such as a \emph{restricted eigenvalue condition} or \emph{compatibility condition}). These assumptions are typically necessary for showing optimal convergence rates \citep{buhlmann2011statistics}
 and also to show correct coverage. They are typically not verifiable in practice. While some results can also be derived under verifiable assumptions \citep{juditsky2011verifiable}, there is still the possibility that the conditions fail to hold. The proposed procedure can in contrast be applied to all design matrices. 

Detecting significant individual variables is typically very difficult in high-dimensional settings due to the presence of clusters of highly correlated variables. The procedure naturally handles confidence intervals for whole groups of variables. 
The lower bound on the confidence interval for the group effect can be computed with convex optimisation (and linear programming in the special case of  $q=1$). All clusters in a hierarchical clustering tree can be  efficiently tested in a top-down approach by starting at the root node and descending into the tree, stopping whenever a cluster of variables is not significant any longer. We have shown that non-trivial bounds can be obtained for groups of highly or even perfectly correlated variables. 

The power of the corresponding testing procedure has been explored empirically. In addition, the theoretical results show that the procedure has high power of detecting important groups of variables as long as the so-called \emph{group effect compatibility condition} is fulfilled, which is a strictly weaker version of the condition necessary to detect the effect of individual variables. If variables are highly correlated within a tested group, the typical assumptions fail to hold, while the \emph{group effect compatibility condition}  is usually still fulfilled. This is also corroborated by the empirical results.  Non-trivial bounds emerge in high-dimensional settings as long as the signal-to-noise ratio is sufficiently large and we test at the right granularity by choosing groups of variables that are large enough to include all highly correlated variables of its members.

\section{Appendix}

\subsection{Additional Lemma}
\begin{lemma}\label{lemma:help}
Let $\mb Y^{(k)} = \mb Y + \mu\cdot \mb E \mbb \gamma^{(k)}$ be the $k=1,\ldots,2m$ vertices, as defined just after~(\ref{eq:bpk}), constructed at  level $\alpha$. With probability at least $1-\gamma$ for $\gamma\in (0,0.2)$, simultaneously  for all $k\in \{1,\ldots,2m\}$, 
\begin{equation}
\label{eq:l11} \| \mb A \mb Y^{(k)} - \mb A E(\mb Y) \|_2^2 \;\le\; 20 \log\big( \frac{1}{\min\{\gamma, \alpha\}}\big) s\frac{\sigma^2}{n} 
\end{equation}
\end{lemma}

\textit{Proof:} 
We can decompose as 
\begin{equation}
\label{eq:l12} \| \mb A \mb Y^{(k)} - \mb A E(\mb Y) \|_2^2 \;\le\; 2\|\mb A \mb Y^{(k)}-\mb A\mb Y\|_2^2 + 2\| \mb A \mb Y -  \mb A E(\mb Y)  \|_2^2.
\end{equation}
The second term on the right hand side of~(\ref{eq:l12}) is equal to $\|\mb A \mbb\varepsilon\|_2^2$. If the errors have a Gaussian distribution, then $\mb A$ will have independent normal entries (using the assumption of an orthonormal $\mb A$) and $(\mb A \mbb\varepsilon)_j \sim\mathcal{N}(0,\sigma^2/n)$ for $j=1,\ldots,s$.
The second term has thus, for Gaussian errors and if divided by $\sigma^2/n$, a $\chi^2_{s}$-distribution. For $\gamma\in (0,0.2)$, the ($1-\gamma$)-quantiles of $Z/s$, where $Z\sim\chi^2_s$, are smaller or equal to  the $(1-\gamma)$-quantiles of a $\chi^2_1$-distributed random variable. For $\gamma\in (0,0.2)$, the ($1-\gamma$)-quantile of 
\[ \frac{\| \mb A \mb Y -  \mb A E(\mb Y)  \|_2^2}{s \sigma^2 /n}\] is thus bounded from above by the corresponding quantile of a $\chi^2_1$-distribution. 
The same is then also true if sub-Gaussian errors are allowed with the appropriate $\sigma^2>0$. Let $q_\gamma$ be the $(1-\gamma)$-quantile of a $\chi_1^2$-distributed random variable.
Using a tail bound for the Gaussian-distribution, \begin{equation}\label{eq:quantboundchi} q_\gamma \le 2\log\big( \frac{2}{\sqrt{2\pi} \gamma}\big) .\end{equation}
Hence, with probability at least $1-\gamma$,
 \[ \| \mb A \mb Y -  \mb A E(\mb Y)  \|_2^2\;\le\; 2\log\big( \frac{2}{\sqrt{2\pi} \gamma}\big) s \frac{\sigma^2}{n} .\]
The first term on the right hand side in~(\ref{eq:l12}), $\|\mb A \mb Y^{(k)}-\mb A\mb Y\|_2^2$, is the distance between the observed response and the vertices, which is exactly the value $\mu$ as per~(\ref{def:Nm}). The radius $\mu$ is chosen as in~(\ref{eq:muC}) as $C>1$ times the ($1-\alpha$)-quantile of $\|\mbb\varepsilon\|_2$, which guarantees that a finite value of the number of vertices for a fixed value of $s$ is sufficient to guarantee the coverage property in~(\ref{def:conhull2}). Thus, using the bound~(\ref{eq:quantboundchi}) on the quantiles of the $\ell_2$-norm of $\|\mb A\mbb\varepsilon\|_2^2$ and the default value $C=3$, we have that
\[ \mu^2=\|\mb A \mb Y^{(k)}-\mb A\mb Y\|_2^2 \;\le\; 9 q_\alpha s\frac{\sigma^2}{n}  \]
and in total we have the left hand side of~(\ref{eq:l12}) is bounded with probability $1-\gamma$ for $\gamma\in (0,0.2)$ by 
\[ \| \mb A \mb Y^{(k)} - \mb A E(\mb Y) \|_2^2 \;\le\; 2(q_\gamma + 9q_\alpha) s\frac{\sigma^2}{n}  \le 20 \max\{q_\alpha,q_\gamma\}s\frac{\sigma^2}{n} \le 20 \log\big( \frac{1}{\min\{\gamma, \alpha\}}\big) s\frac{\sigma^2}{n} ,\]
which completes the proof.
\begin{lemma}\label{lemma:help2}
Let $\mb b^{(k)}$, for $k=1,\ldots,2m$, be the  Basis Pursuit solutions  $\mb b^{(k)} = b(\mb X,\mb Y + \mu\cdot \mb E \mbb \gamma^{(k)})$  at the $2m$ vertices, as defined in~(\ref{eq:bpk}) for level $\alpha$. Under the assumptions of Lemma~\ref{lemma:help} and assumption (A~I), with probability at least $1-\gamma$ for $\gamma\in (0,0.2)$, simultaneously for all $k\in \{1,\ldots,2m\}$,
\[ \| \mb b^{(k)}\|_1 \;\le\; \|\mbb \beta^* \|_1 +  \sqrt{20 \log\big( \frac{1}{\min\{\alpha,\gamma\}} \big)}  \frac{\sigma s }{\varphi_1\sqrt{n} } 
\]
\end{lemma}

\textit{Proof:} By definition~(\ref{eq:bpk}) of $\mb b^{(k)}$,
\begin{equation}\label{eq:bpk2} \mb b^{(k)}\;=\; \mbox{argmin}_{\mb b} \|\mb b\|_1 \mbox{   such that  } \mb A  \mb X \mb b^{(k)} = \mb A \mb Y^{(k)}.\end{equation}
Let $S$ be the set as defined in assumption (A~I). Let $\mb Z$ be the $s\times s$-matrix by keeping all $s$ columns in $\mb A \mb X$ that are in the set $S$. Since $\mb Z$ has full rank by assumption (A~I), $\mb Z^T \mb Z$ is invertible and 
$ \mb Z (\mb Z^T \mb Z)^{-1} \mb Z^T   $
is hence the $s\times s$ identity matrix, so that 
\begin{equation}\label{eq:hl2} \mb Z (\mb Z^T \mb Z)^{-1} \mb Z^T (\mb A \mb Y^{(k)} - \mb A E(\mb Y)) = (\mb A \mb Y^{(k)} - \mb A E(\mb Y)).\end{equation}
Define $\tilde{\mb b}^{(k)}$ by setting 
\begin{align*} \tilde{\mb b}^{(k)}_j & \;=\;   \mbb\beta^*_j \quad \mbox{        if } j\notin S , \\ \mbox{and   } \tilde{\mb b}^{(k)}_S & \;=\; \mbb\beta^*_S + (\mb Z^T \mb Z)^{-1} \mb Z^T (\mb A \mb Y^{(k)} - \mb A E(\mb Y)) .  \end{align*}
Then, using the fact that $\mb A \mb X\mbb \beta^* = E(\mb Y)$, it follows from~(\ref{eq:hl2}) that  
\[ \mb A \mb X \tilde{\mb b}^{(k)} = \mb A \mb Y^{(k)},\]
which means that $\tilde{\mb b}^{(k)}$ is a feasible vector in~(\ref{eq:bpk2}) and hence
\begin{align} \| \mb b^{(k)}\|_1\;&\le\; \| \tilde{\mb b}^{(k)}\|_1 \nonumber \\ \;&\le\;\|\mbb\beta^*\|_1 + \| (\mb Z^T \mb Z)^{-1} \mb Z^T (\mb A \mb Y^{(k)} - \mb A E(\mb Y)) \|_1 \nonumber \\
\;&\le \; \|\mbb\beta^*\|_1 + \sqrt{s} \| (\mb Z^T \mb Z)^{-1} \mb Z^T (\mb A \mb Y^{(k)} - \mb A E(\mb Y)) \|_2 \nonumber \\
\;&\le \; \|\mbb\beta^*\|_1 + (\sqrt{s}/\varphi_1) \|  \mb A \mb Y^{(k)} - \mb A E(\mb Y) \|_2 \label{eq:hh1}
\end{align}
where the last inequality follows since the minimal singular value of $\mb Z$ is larger or equal to $\varphi_1>0$ by assumption (A~I). Now using Lemma~\ref{lemma:help}, with probability at least $1-\gamma$, 
\[ \|  \mb A \mb Y^{(k)} - \mb A E(\mb Y) \|^2_2 \;\le\;  20 s \log\big( \frac{1}{\min\{\alpha,\gamma\}} \big)  \frac{\sigma^2}{n }  ,\]
which, if used in~(\ref{eq:hh1}), completes the proof.

\subsection{Proof of Theorem~\ref{theo:power}}

Recall the definition of the lower bound in $\ell_1$-norm, as defined in~(\ref{def:stat}),
\[ T_G\; :=\; \mbox{min}_{(\mbb \beta,\mbb \eta)\in C_\alpha}  \|\mbb \beta_G\|_1    ,\]
where $C_\alpha$ is replaced in the optimisation with the convex constraint $\bar  C_{m,\mu}$, as defined in~(\ref{def:barC}) with the property that  $C_\alpha \subseteq \bar  C_{m,\mu}$. By definition of $C_{m,\mu}$ as a convex hull over the $k=1,\ldots,2m$ vertices,
\begin{align*}
(\mbb\beta,\mbb \eta) \in \bar   C_{m,\mu} &\Rightarrow \left\{ \begin{array}{rl}  \| \mb A\mb X \mbb \beta - \mb A E(\mb Y)\|_2^2 & \le \max_k \| \mb A\mb Y^{(k)} - \mb A E(\mb Y)\|_2^2 \\ \|\mbb \beta\|_1 &\le \max_k \|\mbb\beta^{(k)}\|_1  \end{array} \right. 
\end{align*} 
 Using Lemma~\ref{lemma:help} and \ref{lemma:help2}, under the made assumption (A~I), with probability at least $1-\gamma$, the right hand sides can be replaced with the relevant uniform bound over all vertices to get 
\begin{align} \label{eq:prb}
(\mbb\beta,\mbb \eta) \in \bar  C_{m,\mu} &\Rightarrow \left\{ \begin{array}{rl}  \| \mb A\mb X \mbb \beta - \mb A E(\mb Y)\|_2^2 & \le       \varphi_1^2\;( \delta\ell)^2 /s    \\ \|\mbb \beta\|_1 &\le \|\mbb\beta^*\|_1 + \delta \ell \end{array} \right. ,
\end{align} 
where $\delta \ell>0$ is given by 
\begin{equation}\label{eq:deltaell} \delta\ell ^2 = 20 \log\big( \frac{1}{\min\{\alpha,\gamma\}} \big) s^2   \frac{\sigma^2 }{\varphi_1^2 n } .\end{equation}
The proof follows by contradiction. Assume there exists a group $G\in \mathcal{G}$ for which for 
\begin{equation}\label{dt}  \delta T \;>\; \max\big\{1, \frac{\varphi_1}{\varphi_2} \sqrt{|S_0|/s} \big\} \, \delta\ell,\end{equation}
the lower bound $T_G$ is too low by at least an amount of $\delta T$,
\begin{equation}\label{eq:asspr} T_G \le \|\mbb\beta^*_G\|_1 - \delta T \end{equation}
We then show that both conditions in~(\ref{eq:prb}) cannot be satisfied simultaneously.

Specifically, we will assume the second condition about the sparsity of the coefficient vector holds in~(\ref{eq:prb}) and show that the first condition in~(\ref{eq:prb}) is then violated. 
Define $\mbb \delta := \mbb \beta - \mbb \beta^*$, where $\mbb \beta$ is the vector for which $T_G=\|\mbb \beta_G\|_1$ and for which there exists a $\mbb \eta$ such that $(\mbb\beta,\mbb \eta) \in \bar   C_{m,\mu} $. 
The strategy is now to show that for all groups $G$ that fulfil the assumptions in the Theorem and for which the lower inequality in~(\ref{eq:prb}) holds, both of the following statements are true 
\begin{align}
\label{I}  (\mathit{I})&:\qquad \|\mbb\delta_{G^c\cap S_0^c}\|_1  \le 2  \big( \|\mbb\delta_{S_0}\|_1-\|\mbb\delta_{G\cap S_0^c} \|_1\big) ,\\
\label{II}  (\mathit{II})&:\qquad \nu_G(\mbb \delta) \ge \delta T.
\end{align}
The inequality~(I) in (\ref{I}) implies via the definition of the \emph{group effect compatibility condition} in~(\ref{eq:scc}) that 
\begin{equation}\label{Ib} \|\mb A \mb X \mbb \delta \|_2^2 \ge \varphi_2^2\; \nu^2_G(\mbb \delta) / |S_0| .\end{equation}
Note that  \[  \mb A \mb X \mbb\delta = \mb A \mb X (\mbb\beta- \mbb\beta^*) = \mb A \mb X \mbb \beta  -  \mb A E(\mb Y),\]
where $\mb Y =\mb X\beta $ as defined above and $\mb X\beta^* = E(\mb Y)$ per definition of $\beta^*$, it follows with~(II) in (\ref{II}) that
\begin{align*} \|\mb A \mb X \mbb \beta -  \mb A E(\mb Y)\|_2^2 \;&\ge \; \varphi_2^2 \;(\delta T)^2 / |S_0| 
\\ \;& >\; \varphi_1^2 (\delta\ell)^2 /s , \end{align*}
where the last inequality follows by~(\ref{dt}). This leads to a contradiction with the first inequality in~(\ref{eq:prb}) and thus proves that for all groups $G\in\mathcal{G}$,
\[ T_G \ge \|\mbb\beta^*_G\|_1 - \delta T,\] with probability at least $1-\gamma$, where $\delta T$ is defined as in~(\ref{eq:asspr}). 
This completes the proof, but it remains to show $(\mathit{I})$ in~(\ref{I}) and $(\mathit{II})$ in~(\ref{II}).

{\bf Proof of (I).} First the proof of $(\mathit{I})$ in (\ref{I}). Since $\mbb \beta^*_{S_0^c}\equiv 0$ by definition of $S_0$ as the set of non-zero coefficients in $\mbb\beta^*$,  and $\mbb \delta = \mbb \beta - \mbb \beta^*$,
\begin{align*} \|\mbb \beta\|_1 & \ge \|\mbb \beta^*_{S_0}\|_1 - \|\mbb \delta_{S_0}\|_1 + \|\mbb\delta_{S_0^c} \|_1 \\ & =\|\mbb \beta^*_{S_0}\|_1 - \|\mbb \delta_{S_0}\|_1 + \|\mbb\delta_{G\cap S_0^c} \|_1 + \|\mbb\delta_{G^c\cap S_0^c} \|_1 ,\end{align*}
Combining with the assumed second condition in~(\ref{eq:prb}) ($\|\mbb \beta\|_1 \le \|\mbb\beta^*\|_1 + \delta \ell$) and using $\|\mbb \beta^*_{S_0}\|_1 = \|\mbb \beta^*\|_1$,
\begin{align} \|\mbb\delta_{G^c\cap S_0^c}\|_1  &\le  \|\mbb\delta_{S_0}\|_1-\|\mbb\delta_{G\cap S_0^c} \|_1  +\delta\ell \nonumber \\ &= \big( \|\mbb\delta_{S_0}\|_1-\|\mbb\delta_{G\cap S_0^c} \|_1\big) \big(1  + \frac{\delta\ell}{\|\mbb\delta_{S_0}\|_1-\|\mbb\delta_{G\cap S_0^c} \|_1}). \label{eq:hpa}\end{align}
The assumption~(\ref{eq:asspr}) together with~(\ref{dt}) implies $\|\mbb \beta_G\|_1 \le \|\mbb \beta^*_G\|_1 - \delta\ell$. Since  also
\begin{equation}\label{l1b} \|\mbb \beta_G\|_1 \ge \|\mbb \beta^*_{G\cap S_0}\|_1 - \|\mbb \delta_{G\cap S_0}\|_1 + \|\mbb \delta_{G \cap S_0^c}\|_1  , \end{equation} it follows with $\|\mbb \beta^*_{G\cap S_0}\|_1= \|\mbb\beta^*_{G}\|_1$ that
\[ \|\mbb\delta_{G\cap S_0} \|_1 \ge \delta\ell + \|\mbb \delta_{G\cap S_0^c}\|_1 ,\]
and thus also $\|\mbb \delta_{S_0}\|_1 \ge \delta\ell + \|\mbb \delta_{G\cap S_0^c}\|_1 $.
The factor on the right hand side of~(\ref{eq:hpa}) is thus bounded by 
\[   \big(1  + \frac{\delta\ell}{\|\mbb\delta_{S_0}\|_1-\|\mbb\delta_{G\cap S_0^c} \|_1}) \;\le\;   2 \]
Using this in~(\ref{eq:hpa}), we get the inequality
\[  \|\mbb\delta_{G^c\cap S_0^c}\|_1  \le 2  \big( \|\mbb\delta_{S_0}\|_1-\|\mbb\delta_{G\cap S_0^c} \|_1\big) ,\]
which shows that $(\mathit{I})$ in~(\ref{I}) is true. 

{\bf Proof of (II).} It remains to show $(\mathit{II})$ in~(\ref{II}). 
A refinement of~(\ref{l1b}) yields
\begin{align*} \|\mbb \beta_G\|_1 &=\; \sum_{k\in G} |\beta_k|\; =\; \sum_{k\in G} |\beta^*_k+\delta_k| \\ 
& \ge \sum_{k\in G \cap S_0} \Big( |\beta_k^*| - \mbox{sign}(\beta^*_k) \delta_k \Big) + \sum_{k\in G \cap S_0^c} |\delta_k|\\
& \ge  \sum_{k\in G \cap S_0} |\beta_k^*| - \min_{\varsigma \in \mathcal{S}}  \sum_{k\in G} \varsigma_k \delta_k ,
\end{align*}
where $\mathcal{S}\subseteq[-1,1]^p$ is per Definition~\ref{def:gcc} the subgradient of the $\ell_1$-norm evaluated at $\mbb \beta^*$.
Thus 
\begin{align*}  \|\mbb \beta_G\|_1 & \ge \|\mbb\beta^*_{S_0\cap G}\|_1 - \nu_G(\mbb \delta) \; = \|\mbb\beta^*_{G}\|_1 - \nu_G(\mbb \delta),\end{align*}
having used the definition of $\nu_G(\mbb \delta)$ in~(\ref{eq:nu}). Since assumption~(\ref{eq:asspr}) implies again
$\|\mbb \beta_G\|_1 \le \|\mbb \beta^*_G\|_1 - \delta T  $,
it follows that $ \nu_G(\mbb \delta)\ge \delta T  $, which completes the proof of $(\mathit{II}) $ in~(\ref{II}). Since we have now shown 
$(\mathit{I}) $ and $(\mathit{II}) $, the proof of the theorem is complete.

\bibliographystyle{plainnat}
\bibliography{text3.bbl}

\end{document}